\documentclass[twocolumn,numberedappendix]{emulateapj}
\bibpunct[; ]{(}{)}{;}{a}{}{;}
\usepackage{apjfonts}

\usepackage{graphicx}
\usepackage{natbib}

\newcommand{\Msolar}{M${_\odot}$\,}

\shorttitle{Recent star formation in 30 Dor}
\shortauthors{De Marchi, Panagia \& Beccari}

\journalinfo{The Astrophysical Journal}
\received{10 May 2017}
\accepted{9 Aug 2017}

\begin{document}

\title{Photometric determination of the mass accretion rates of 
pre-main sequence stars. \\ V. Recent star formation in the 30 Dor
Nebula\altaffilmark{*}}

\author{
Guido De Marchi,\altaffilmark{1}
Nino Panagia,\altaffilmark{2,3,4}
and Giacomo Beccari\,\altaffilmark{5}
}

\altaffiltext{1}{European Space Research and Technology Centre,
Keplerlaan 1, 2200 AG Noordwijk, Netherlands; gdemarchi@esa.int}
\altaffiltext{2}{Space Telescope Science Institute, 3700 San Martin
Drive, Baltimore MD 21218, USA; panagia@stsci.edu}

\altaffiltext{3}{INAF--NA, Osservatorio Astronomico di Capodimonte,
Salita Moiariello 16, 80131 Napoli, Italy}
\altaffiltext{4}{Supernova Limited, OYV \#131, Northsound Rd., Virgin Gorda 
VG1150, Virgin Islands, UK}
\altaffiltext{5}{European Southern Observatory, Karl--Schwarzschild-Str.
2, 85748 Garching, Germany; gbeccari@eso.org}

\altaffiltext{{$\star$}}{Based on observations with the NASA/ESA
{\it Hubble Space Telescope}, obtained at the Space Telescope Science
Institute, which is operated by AURA, Inc., under NASA contract
NAS5-26555}

\begin{abstract}  

We report on the properties of the low-mass stars that recently formed
in the central $\sim 2\farcm7 \times 2\farcm7$ of 30\,Dor including the
R\,136 cluster. Using the photometric catalogue of De Marchi et al.
(2011c), based on observations with the {\em Hubble Space Telescope}
(HST), and the most recent extinction law for this field, we identify
$1\,035$ bona-fide pre-main sequence (PMS) stars showing $H\alpha$
excess emission at the $4\,\sigma$ level with $H\alpha$ equivalent width
of 20\,\AA\, or more. We find a wide spread in age spanning the range
$\sim 0.1 - 50$\,Myr. We also find that the older PMS objects are placed
in front of the R\,136 cluster and are separated from it by a
conspicuous amount of absorbing material, indicating that star formation
has proceeded from the periphery into the interior of the region. We
derive physical parameters for all PMS stars, including masses $m$, ages
$t$, and mass accretion rates $\dot M_{\rm acc}$. To identify reliable
correlations between these parameters, which are intertwined, we use a
multivariate linear regression fit of the type $\log \dot M_{\rm acc} =
a \times \log t +b \times \log m +c$. The values of $a$ and $b$ for 
30\,Dor are compatible with those found in NGC\,346 and NGC\,602. We
extend the fit to a uniform sample of $1\,307$ PMS stars with $0.5 <
m$/\Msolar$ < 1.5$ and $t<16$\,Myr in six star forming regions in the
Large and Small Magellanic Clouds and Milky Way with metallicities in
the range $0.1 - 1.0$\,Z$_\odot$. We find $a=-0.59 \pm 0.02$ and $b=0.78
\pm 0.08$. The residuals are systematically different between the six
regions and reveal a strong correlation with metallicity $Z$, of the
type $c=(-3.69\pm0.02)- (0.30\pm0.04) \times \log Z/Z_\odot$. A possible
interpretation of this trend is that when the metallicity is higher so
is the radiation pressure and this limits the accretion process, both in
its rate and duration.

\end{abstract}


\section{Introduction}

The Tarantula Nebula (30\,Dor) in the Large Magellanic Cloud (LMC) is
the nearest extragalactic region of extended star formation. It is
energized by the OB association NGC\,2070, which hosts at its centre the
Radcliffe\,136 (R\,136) cluster, the closest example of a massive
extragalactic starburst region (Walborn 1991). Ground-based observations
dating back to the 1980s already revealed several discrete epochs of
formation of massive stars within the nebula (e.g. Walborn 1986, 1991;
Melnick 1987), but only the advent of the refurbished {\em Hubble Space
Telescope} (HST) allowed an in-depth study also of lower mass stars
(e.g. Sirianni et al. 2000; Zinnecker et al. 2002; Andersen et al.
2009). 

De Marchi et al. (2011c) presented a study of the populations of
low-mass stars  recently formed in the central regions of the 30\,Dor
nebula, using observations with the {\em Wide Field Camera\,3} (WFC\,3)
on board the HST. Looking for objects with $H\alpha$ excess emission,
these authors discovered a population of more than 1\,150 pre-main
sequence (PMS) stars in a region of $\sim 40 \times 40$\,pc$^2$ in
NGC\,2070, including the central R\,136 cluster. Although many of these
PMS stars have young ages ($< 4$\,Myr) and a spatial distribution
matching that of the massive members of R\,136, the study of De Marchi
et al. (2011c) revealed that about one third of the PMS objects have
ages in excess of $\sim 10$\,Myr and are preferentially located to the
east of R\,136. This indicates that formation of low-mass stars in this
field has proceeded over a period of time longer than that associated
with the birth of the massive stars of R\,136 some $\sim 2-3$\,Myr ago
(e.g. De Marchi et al. 1993; Hunter et al. 1995a; Massey \& Hunter 1998;
de Koter, Heap \& Hubeny 1998; Walborn et al. 1999).

The presence in the Tarantula nebula of stars older than the massive
members of R\,136 was already established earlier. For instance, a
spectroscopic study of massive ($ > 40$\,\Msolar) stars conducted by
Selman et al. (1999) allowed them to identify three bursts of increasing
strength, occurring respectively 5, $2.5$, and less than $1.5$\,Myr ago.
This had been noticed before by Brandl et al. (1996). Comparing their
multi-band $UVIHK$  photometry with evolutionary tracks (Meynet et al.
1994), Brandl et al. (1996) determined the age of stars more massive
than 12\,\Msolar within a field of $\sim 10$\,pc$^2$ around R\,136,
detecting three peaks of massive star formation separated by
approximately $2.5$\,Myr. Similarly, from a morphological study of the
ages and spatial distribution of OB stars, Walborn \& Blades (1997) had
identified multiple generations of stars in 30\,Dor, with ages of $<1$,
2--3, and 6\,Myr inside R\,136. Furthermore, for the Hodge\,301 cluster,
located $3^\prime$  north-west of it, they obtained an age of $\sim
10$\,Myr or older, as already suggested by Lortet \& Testor (1991). An
even older age for Hodge\,301 was later derived by Grebel \& Chu (2000).
Using both photometry and spectroscopy, these authors concluded that the
most likely age for Hodge\,301 is of the order of 20 -- 25\,Myr, as at
younger ages isochrones would not match both the main-sequence (MS)
turn-off and the locus of the blue and red supergiants. From the colour
and magnitude of the PMS sequence turn-on, where the PMS joins the MS,
Cignoni et al. (2016) recently concluded that Hodge 301 is older still,
with ages in the range $26.5 - 31.5$\,Myr.

In general, these conclusions on the age of the populations in 30\,Dor
are based on the properties of the relatively short-lived massive stars.
It is thus difficult to establish whether the massive stars in R\,136
and the less recent episodes around 30\,Dor are related or simply happen
to have occurred in neighbouring regions. On the other hand, the PMS
stars identified by De Marchi et al. (2011c) in 30\,Dor have masses in
the range $0.5 - 3$\,\Msolar and as such they can be used to trace the
progression of star formation on scales of several 10\,Myr. For
instance, a typical $\sim 1$\,\Msolar PMS star of metallicity $Z=0.007$,
as appropriate for the LMC, takes about 25\,Myr to reach the MS (e.g.
Tognelli et al. 2012). Throughout this phase, objects can be identified
as PMS stars thanks to the emission features associated with the mass
accretion process and through their physical parameters. Thus their mass
accretion rate and age can be derived for a statistically significant
number of stars, typically several hundreds. 

In this work, we follow up on the original study of De Marchi et al.
(2011c), starting from the same observational material, yet updated to 
the most recent WFC\,3 calibration (Deustua et al. 2017).
Furthermore, we take advantage of the accurate 30\,Dor extinction
properties in the range $0.3 - 1.6\,\mu$m as derived by De Marchi \&
Panagia (2014; see also De Marchi et al. 2016) from the colours and
magnitudes of about 180 red giant stars uniformly distributed over this
field. This will allow us to obtain accurate physical parameters for all
PMS stars in this region, including their mass accretion rates. Studies
conducted so far in the Magellanic Clouds (MCs) have revealed that the
accretion  process appears to be more intense and to last longer than in
nearby Galactic star forming regions (De Marchi, Panagia \& Romaniello
2010; De Marchi et al. 2011a; De Marchi, Panagia \& Sabbi 2011; Spezzi
et al. 2012; De Marchi, Beccari \& Panagia 2013).  

{It has been reported in the literature (e.g. Yasui et al. 2009,
2010, 2016) that in some star-forming regions of low metallicity in the
extreme outer Galaxy the disc lifetime appears to be shorter than that
measured in similar regions at solar metallicity. However, these
conclusions are based on observations of the near-infrared (NIR)
emission from the dust component in the discs, using dust as a proxy for
the total mass of the discs. Therefore, these results are not directly
comparable with our measurements of the gas accretion process in
low-metallicity environments. In fact, optical spectroscopic
observations of similar small low-metallicity star-forming regions in
the Galactic anticentre (Cusano et al. 2011; Kalari \& Vink 2015)
indicate that a significant fraction of the young stellar objects have
actually preserved their discs and envelopes and continue to accrete.
And indeed, if the metallicity is lower also the dust abundance is
lower, and a proportionally lower amount of stellar radiation is
absorbed by grains. Therefore, for an equal stellar luminosity the disc
NIR emission is also lower. Thus, the NIR detectability of a low
metallicity disc appears to drop faster than it is found for higher
metallicity sources. For this reason we favour studying the time
evolution of discs focusing on their gaseous component, which represents
the bulk of the circumstellar disc mass.}


The ultimate goal of this investigation is to explore how the accretion
process during the PMS phase depends on the mass of the forming star and
how it is affected by the chemical composition of the parent molecular
cloud. The mini-starburst nature of 30\,Dor (Leitherer 1998) makes this
region crucial for our understanding of star formation also in a
cosmological context, since it offers an environment with conditions
similar to those prevailing at redshift  $z \simeq 2$, when star
formation in the Universe was at its peak (e.g. Madau et al. 1996; Lilly
et al. 1996). 

The structure of the paper is as follows. The observations and
photometric catalogue are briefly summarised in Section\,2. In
Section\,3 we deal with the correction for extinction. Section\,4 is
devoted to the identification of the PMS stars and to their physical
properties. In Section\,5 we study how the mass accretion rate of PMS
stars is related with stellar parameters. A comparison with other
star-forming regions is provided in Section\,6. A summary of the most
important conclusions of the paper is presented in Section\,7. 


\section{Observations and photometry}

The data used in this work are extracted from the multiband photometric
catalogue presented by De Marchi et al. (2011c), based on observations
obtained with the WFC\,3 on board the HST of a region of $\sim
160\arcsec \times 160\arcsec$ at the centre of NGC\,2070, including the
R\,136 cluster. The observations were collected in several bands over
the range $0.3-1.6\, \mu$m, namely through the F336W, F438W, F555W,
F656N, F110W and F160W filters. In this work we will concentrate on the
observations in the F555W, F814W, and F656N bands (hereafter $V$, $I$, 
and $H\alpha$, respectively). Details on data reduction and photometry
are given in De Marchi et al. (2011c). {With respect to the original
catalogue, we updated the zero points of the photometric calibration in
the F555W, F656N, and F814W bands according to the most recent values
(Deustua et al. 2017).}

\begin{figure}[t]
\centering
\resizebox{\hsize}{!}{\includegraphics{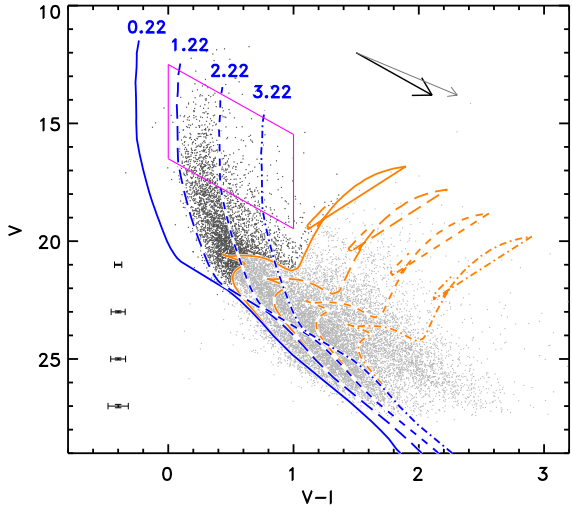}} 
\caption{CMD containing all $\sim 18\,800$ stars with photometric
uncertainty of less than $0.1$\,mag in $V$ and $I$. Typical
uncertainties on colour and magnitude are shown for stars with $V>20$.
The arrows indicate the reddening vectors applicable to the 30\,Dor
nebula (thick arrow) and MW (thin arrow) for $A_V=1.8$. Theoretical
isochrones from Marigo et al. (2008) are also shown, for ages of 3\,Myr
(blue) and 2\,Gyr (orange), metallicity $Z=0.007$ and a distance modulus
$(m-M)_0=18.55$. Solid lines include only the MW intervening absorption
along the line of sight ($A_V=0.22$, $E(V-I)=0.1$). The other line types
are further reddened for purpose of illustration by respectively $\Delta
A_V=$1, 2, and 3 magnitudes (as indicated) according to the 30\,Dor 
extinction law. The box contains reference stars in the upper main
sequence used for reddening determination. The slanted sides of the box
are parallel to the 30\,Dor reddening vector (thick arrow). } 
\label{fig1}
\end{figure}

A total of about 18\,800 sources are detected simultaneously in the $V$
and $I$ bands with photometric uncertainty of less than $0.1$\,mag in
both bands. The corresponding colour--magnitude diagram (CMD) is shown
in Figure\,\ref{fig1}. In Sections\,3 and 4 we will use this CMD, the
photometry in the $H\alpha$ band, and the spatial distribution of the
stars in the field to identify PMS objects and correct their photometry
for the effects of differential reddening. We will adopt the method
already employed by De Marchi et al. (2011c), briefly summarised in the
following section, and refer the reader to that work for further
details.

\section{Differential extinction}

The upper MS (UMS) seen in Figure\,\ref{fig1} at $V \lesssim 20$ is
considerably wider than the photometric uncertainty (see the error bars
in the figure), as De Marchi et al. (2011c) already pointed out. This is
due to the large amount of differential extinction known to affect this
field (Fitzpatrick \& Savage 1984; Ma{\'\i}z Apell{\'a}niz et al. 2014).
De Marchi \& Panagia (2014) measured the extinction towards about 180
red giants in this field and derived the extinction law in the range
$0.3-1.6\,\mu$m. They found that at optical wavelengths the extinction
curve is almost parallel to that of the diffuse Galactic interstellar
medium, and the presence of an extra grey component due to a larger
fraction of large grains results in a higher value of the ratio of total
and selective extinction  $R_V \equiv A_V/E(B - V) = 4.5 \pm 0.2$. The
corresponding reddening vector is shown by the thick arrow in
Figure\,\ref{fig1}. For comparison, we also show as a thin arrow the
reddening vector corresponding to $R_V=3.1$ that was originally adopted
by De Marchi et  al. (2011c). The steeper reddening vector corresponding
to the extinction law measured by De Marchi \& Panagia (2014; see also
De Marchi et al. 2016) has important implications for the stellar
parameters derived from the photometry, also for PMS stars, as it
typically results in higher stellar masses. 

Although the extinction law is now known in this field, the reddening to
each object still has to be determined individually, due to the patchy
nature of the absorption. 

The solid blue line in Figure\,\ref{fig1} corresponds to a theoretical
isochrone from Marigo et al. (2008) for an age of 3\,Myr\footnote{For
the UMS stars in the range of interest here (i.e. in the observed $V$
range $12 - 16$ corresponding to masses in the range about
$15-50$\,\Msolar), there is no noticeable difference in the CMD between
the Padova post-MS models (Marigo et al. 2008) and pre-MS models (e.g.
Bressan et al. 2012) for an age of 3 Myr}. This is the accepted age for
the massive members of R\,136  (e.g. De Marchi et al. 1993; Hunter et
al. 1995), even though they might be slightly younger ($\sim 2$\,Myr) if
Wolf Rayet stars are still burning hydrogen (Massey \& Hunter 1998; de
Koter, Heap \& Hubeny 1998; Walborn et al. 1999). We have adopted a
distance of $51.4 \pm 1.2$\,kpc (Panagia et al. 1991; Panagia 1999),
corresponding to a distance modulus $(m-M)_0=18.55$, and a metallicity
$Z=0.007$ since this is a typical value for the LMC (e.g. Hill,
Andrievsky \& Spite 1995; Geha et al. 1998). The isochrone also includes
the effects of the Milky Way (MW) intervening absorption along the line
of sight. According to Fitzpatrick \& Savage (1984), this amounts to
$A_V=0.22$ and $E(V-I)=0.1$. The long-dashed, short-dashed, and
dot-dashed blue lines are the same isochrone further reddened for
purpose of illustration by respectively 1, 2, and 3 magnitudes in $A_V$
according to the 30\,Dor extinction law (De Marchi \& Panagia 2014; De
Marchi et al. 2016), corresponding to an additional contribution of
$0.34$, $0.68$ and $1.02$\,mag to the $E(V-I)$ value.  

These isochrones highlight the reddening spread amongst the young stars
in the UMS. The situation is further complicated by the presence of the
older LMC field population. {Both ground-based and HST observations 
concur in assigning ages of $1-3$\,Gyr to the intermediate-age
populations in clusters and in the field of the LMC (e.g. Westerlund,
Linde \& Lyng\aa 1995; Elson, Gilmore \& Santiago 1997; Geha et al. 1998),
with a major star-formation event occurring about 2\,Gyr ago.
Therefore,} for comparison, the orange lines show isochrones for an age
of 2\,Gyr and the same metallicity, distance, and reddening values used
for the 3\,Myr isochrones, also from the models of Marigo et al. (2008).
They reveal a different and even wider reddening spread for the older
population (see De Marchi \& Panagia 2014 and De Marchi et al. 2016 for
details). They also reveal that stars of different ages and with
significantly different extinction can occupy the same region of the
CMD. For instance, most stars with $V-I > 0.5$ and $V > 20$ could be
both PMS objects, or red giants, or even older MS stars. This is why our
search for PMS objects will rely on $H\alpha$ excess emission as a
signature of youth.

In such a complex field, a reddening correction can be derived only for
objects whose un-extinguished location in the CMD is known. These
include, for instance, red giant stars belonging to the red clump (RC,
see De Marchi, Panagia \& Girardi 2013; De Marchi \& Panagia 2014; De
Marchi et al. 2016). The same applies to UMS stars. The box marked in 
Figure\,\ref{fig1} contains $\sim 700$ objects whose reddening is simply
given by the displacement from the theoretical isochrone for the
appropriate age, metallicity and distance. These stars serve as a
reference to derive a statistical reddening correction for low-mass
stars with a similar age and spatial distribution. This can be done with
the ``nearest neighbours'' method, introduced by Romaniello et al.
(2002) and described in detail in De Marchi et al. (2011a, 2011c). In
short, it consists in assigning to the identified PMS stars a reddening
value determined as the weighted average of those of the reference UMS
stars in their vicinity. This procedure implicitly assumes that stars
projected within a relatively small region of the field have similar
extinction values. This has been verified using the UMS stars
themselves, which as a test were de-reddened in the same way using the
other reference UMS stars. This  resulted in a reduced colour spread of
the UMS in the CMD (see De Marchi et al. 2011a, 2011c for details). The
nearest neighbour correction will be used for the younger PMS stars
identified in Section\,4.

{One should note that there is an additional source of reddening
that we are not explicitly considering. If the circumstellar disc has a 
high inclination angle, it can intercept the light of the star, causing
further extinction. This event, however, has a low probability of
happening. De Marchi et al. (2013b) conducted a detailed study of the
effects of disc inclination on the apparent colour and magnitude of PMS
stars in NGC\,6611. They show that the effect starts to be noticeable
when the inclination angle reaches $\sim 80^\circ$, thus in principle
affecting 15\,\% of the objects. However, when the angle exceeds
$85^\circ$ the central star becomes typically 4 to 5 mag fainter,
thereby becoming undetectable in most cases. Thus, one should expect
that less than 10\,\% of the stars in our sample may have extra
reddening caused by the circumstellar disc. Therefore the effect is
negligible.} 

\section{Searching for PMS stars}

As mentioned above, differential reddening makes it impossible to identify 
bona-fide PMS stars based on their broad-band photometry alone. Instead
we search for stars with a strong $H\alpha$ excess emission. This
feature is characteristic of low-mass star formation and is attributed
to the gravitational energy released by in-falling matter that ionises
and excites the surrounding gas  (e.g. K\"onigl 1991; Shu et al.
1994). A reliable method to detect these objects using
multi-colour photometry is addressed in a series of papers by De Marchi
et al. (2010, 2011a, 2011b, 2011c, 2013a), to which we refer the reader. 

Briefly, the median $V-I$ and $V - H\alpha$ colours of stars with small
($< 0.05$\,mag) photometric uncertainties in the $V, I$ and $H\alpha$
bands define an empirical reference template against which we identify
objects with excess $H\alpha$ emission. Even before correction for
reddening, the colour--colour diagram $V-H\alpha$, $V-I$ provides a
robust identification of stars with $H\alpha$ excess, since in these
bands the reddening vector runs almost parallel to the median
photospheric colours of normal stars (De Marchi et al. 2010). 

\begin{figure}[t]
\centering
\resizebox{\hsize}{!}{\includegraphics{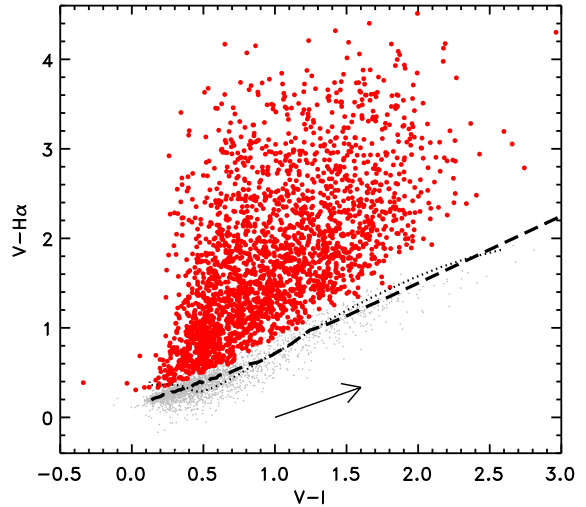}}
\caption{Identification of objects with excess $H\alpha$ emission. The
dashed line is the median photospheric $V-H\alpha$ colour for stars with
small ($<0.05$\,mag) combined photometric uncertainties in all three
bands. {For comparison, the dotted line shows the colours in these
filter for the model atmospheres of Bessell et al. (1998).} Thick dots
denote stars with $H\alpha$ excess emission higher than the $3\,\sigma$
level. The $A_V=1.8$ reddening vector is shown, as an example, for the
30\,Dor extinction law.} 
\label{fig2}
\end{figure}

Following this procedure, we identified stars with $V-H\alpha$ colour
exceeding the $V-H\alpha$ colour of normal stars with no excess and with
the same $V-I$ by at least $0.1$\,mag or by three times the photometric
uncertainty on the $V-H\alpha$ colour (i.e. $3\,\sigma$), whichever is
larger. We further restricted the search to stars with combined
photometric uncertainty not exceeding $0.15$\,mag in all three bands and
not larger than $0.1$\,mag in $V$ and $I$ individually. This results in
about  $2\,175$ PMS candidates, shown as thick dots in
Figure\,\ref{fig2}. 

The dashed line in the figure represents the median photospheric
$V-H\alpha$ colour as a function of $V-I$ for stars with small
photometric uncertainties in all three bands. These are typically older
MS stars and do not have appreciable $H\alpha$ excess. For comparison,
the thin dotted line shows the model atmospheres of Bessell, Castelli \&
Pletz (1998) for stars with effective temperatures in the  range 3\,500
\,K $\le T_{\rm eff} \le $ 40\,000\,K, surface gravity $\log g=4.5$ and
a metallicity index $[M/H]=-0.5$, as appropriate for the LMC  (Dufour
1984). The theoretical colours, calculated for the specific WFC\,3
filters, already include a typical reddening value of $E(V-I)=0.5$, made
up of $E(V-I)=0.1$ due to intervening MW absorption along the
line of sight and $E(V-I)=0.4$ due to LMC absorption in front of and
within 30\,Dor. As we will see later, this is a representative average
reddening value for this field.


\begin{figure}[t]
\centering
\resizebox{\hsize}{!}{\includegraphics{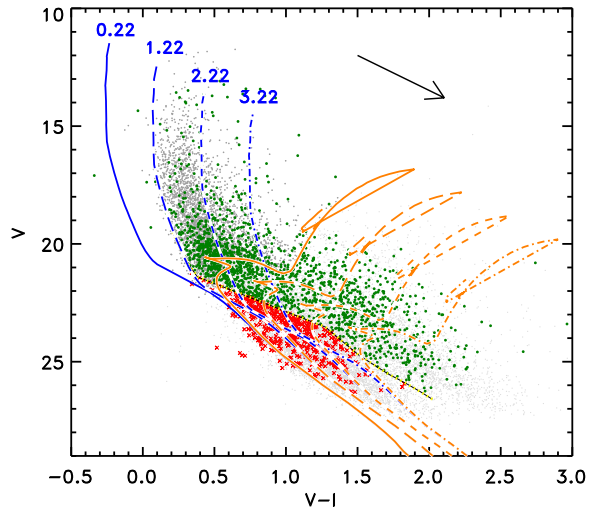}}
\caption{Same as Figure\,\ref{fig1}, with in addition all candidate PMS
stars with $H\alpha$ excess emission and the isochrone for PMS age of
5\,Myr (thin dashed line) for $A_V=1.22$. Stars with $H\alpha$ excess
emission located above and below the 5\,Myr isochrone are indicated
respectively with green thick dots and red crosses and are approximately 
younger and older than 5\,Myr.} 
\label{fig3}
\end{figure}

Objects with $H\alpha$ excess emission are shown as thick symbols in
Figure\,\ref{fig3}. Following De Marchi et al. (2011c), we will consider
their positions in the CMD in relation with their spatial distribution
in order to determine how to correct their photometry for reddening
using neighbouring reference UMS stars. All these objects with
$H\alpha$  excess emission are candidate PMS stars, and as such are
relatively young. We used a theoretical PMS isochrone for a rough
preliminary separation of younger and older PMS stars. A more accurate
age determination will be possible after correction for reddening. 

The thin short-dashed line is the theoretical isochrone for a PMS age of
5\,Myr and metallicity $Z=0.007$ from the models of Tognelli et al.
(2012), translated into the WFC\,3 bands for the assumed distance
modulus and a combined reddening $A_V=1.22$. Using the 5\,Myr PMS
isochrone as a reference, we have marked as thick dots and crosses
respectively the objects above and below that line, and hence
approximately younger and older than 5\,Myr.

\begin{figure}[t]
\centering
\resizebox{\hsize}{!}{\includegraphics{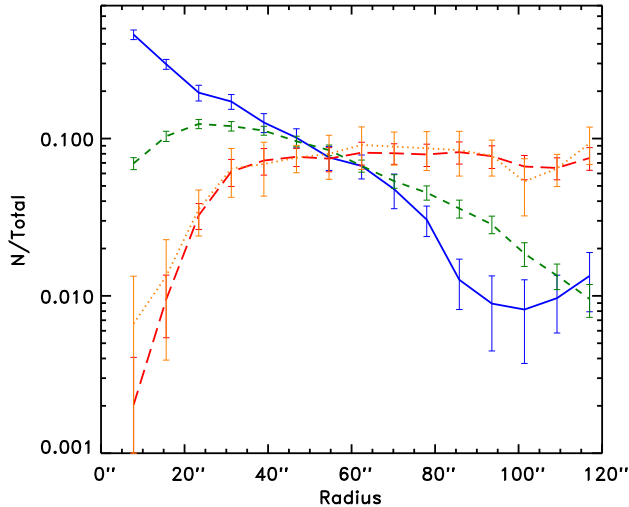}}
\caption{Radial distribution of UMS stars (solid line, blue), and of 
low-mass PMS candidates younger than 5\,Myr (short-dashed line, green) 
and older than 5\,Myr (long-dashed line, red). The dotted line
corresponds to stars with ages intermediate between 5 and 10\,Myr. The
curves are  normalised by the number of objects in each sample 
and the solid line is further rescaled by a factor of $0.65$ for ease of
comparison. Note that crowding is the cause of the apparent drop
observed at small radii {in the profiles of PMS stars (the effect is
negligible for the much brighter and sparser UMS objects)}. Error bars
correspond to the Poisson uncertainty on the number counts.}
\label{fig4}
\end{figure}

The radial distributions of these objects are shown in
Figure\,\ref{fig4}, drawn from the nominal centre of R\,136. De Marchi
et al. (2011c) had already shown that stars younger than 5\,Myr
(short-dashed line) follow  remarkably well the distribution of
reference UMS stars (solid line, rescaled by a factor of $0.65$ to ease
the comparison), while stars older than 5\,Myr (long-dashed line) have a
rather uniform distribution. The dotted line shows for comparison the
distribution of stars between the 5\,Myr and 10\,Myr isochrones {(the
latter is not shown in Figure\,\ref{fig3} but is also from the models of
Tognelli et al. 2011). The dotted line} confirms that even the least old
PMS candidates have a radial distribution different from that of UMS
stars. Thus, we can safely use the nearby reference UMS stars to derive
a reddening correction also for similarly young PMS candidates, but not
for the older ones.  

We assigned to each young PMS candidate a reddening value determined as
the weighted average of the reddenings of the five closest reference UMS
stars within 5\,pc (20$\arcsec$), using as weight $1/d^2$ where $d$ is
the projected distance. All young PMS candidates have at least 2
reference UMS neighbours within 5\,pc with a median number of 15. The
reddening values for young PMS candidates vary from $E(V-I)=0.57$ to
$0.89$, or $A_V=1.63$ to $A_V=2.59$ (respectively the 17 and 83
percentiles), with a median $E(V-I)=0.71$ or $A_V=2.05$. All these
values include the $A_V=0.22$ Galactic component along the line of
sight. 

To determine the uncertainty introduced by this reddening correction, we
applied the same method to UMS stars, using as neighbours the five
closest reference UMS objects within 5\,pc, excluding the star itself.
Compared to the $V-I$ colour that the star would have when translated
along the reddening vector to the 3\,Myr isochrone (solid blue line in
Figure\,\ref{fig1}), we found a typical difference of $0.1$\,mag
($1\,\sigma$). We will use this value as a measure of the statistical
uncertainty introduced by the nearest neighbour reddening
correction.    

For the older PMS candidates using the nearest neighbours would not be
appropriate, owing to their different spatial distribution (see
Figure\,\ref{fig4}). In this case, the only acceptable solution is to
apply to all older PMS stars one and the same reddening value. The
minimum possible value is $A_V=0.22$, which corresponds to the Galactic
intervening extinction. The maximum statistically acceptable value is
one that would move no more than $\sim 1/6$ of the stars to the left of
the adopted MS, i.e. that would make them appear bluer than the solid
blue line in Figure\,\ref{fig3}. This would correspond to the
$1\,\sigma$ deviation. This is the case for $A_V=1.22$. Thus, all
reddening values in the range $0.22 < A_V < 1.22$ are possible. We
will adopt $A_V=0.72 \pm 0.50$ as a representative value and will
specifically address the effects that assuming $A_V=0.22$ or $A_V=1.22$
for these candidate PMS stars would have on the derived physical
parameters. 

An important conclusion that we can already draw is that the older PMS
candidates are placed in front of the R\,136 cluster and are separated
from it by a conspicuous amount of absorbing material. This stems from
the fact that: 1) the maximum reddening towards these objects
($A_V=1.22$) is considerably lower than the reddening towards the
younger PMS stars (17 percentile $A_V=1.63$); and 2) their spatial
distribution is uniform over the field. The separation in $A_V$ and
spatial distribution confirms that younger and older low-mass PMS
candidates are associated with distinct star formation episodes across
the volume probed by these observations. In particular, it appears that
star formation has proceeded into the region.  

\begin{figure}[t]
\centering
\resizebox{0.89\hsize}{!}{\includegraphics{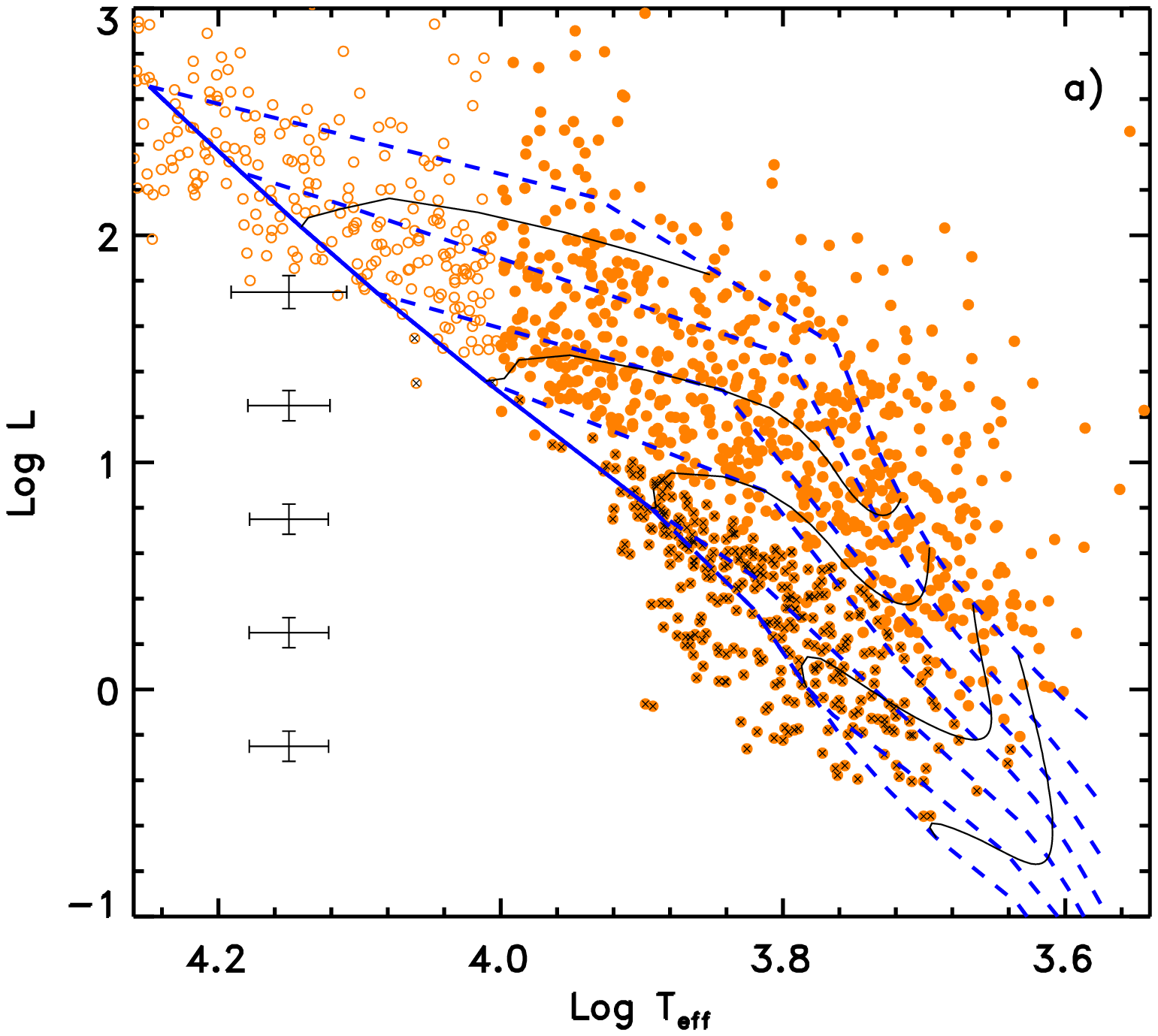}}
\resizebox{0.89\hsize}{!}{\includegraphics{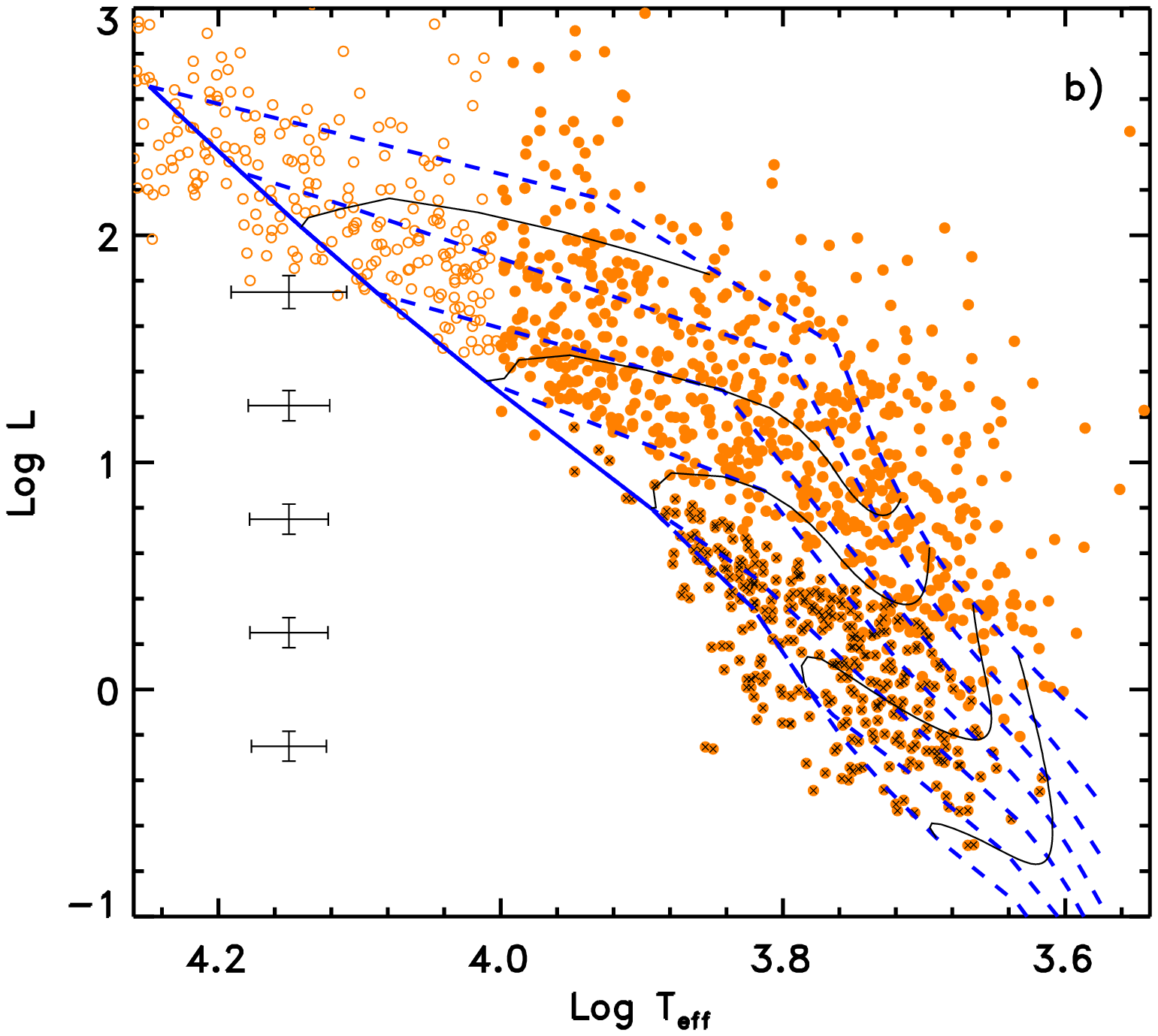}}
\resizebox{0.89\hsize}{!}{\includegraphics{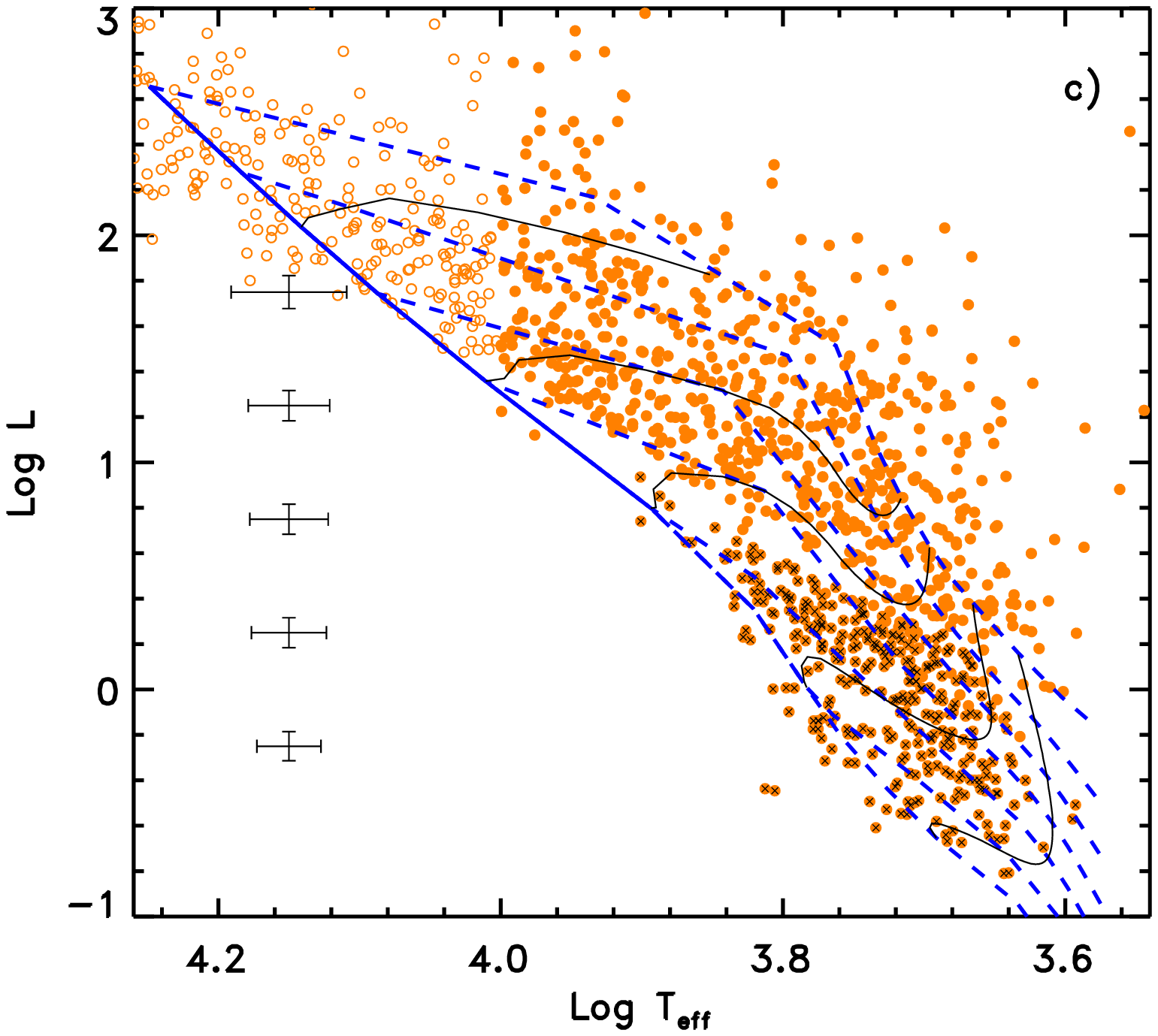}}
\caption{H--R diagram of the bona-fide PMS stars (filled dots, with an
additional cross if older than 5\,Myr). Objects hotter than 10\,000\,K
(unfilled dots) are shown but not not considered in this work. The
typical $\pm 1\,\sigma$ uncertainties on $T_{\rm eff}$ and $L$ are shown
by the error bars. The three panels correspond to different reddening
values for the stars classified as older than 5\,Myr in the CMD of
Figure\,\ref{fig3} (crosses). The adopted values are, respectively,
$A_V=1.22$ (panel a), $0.72$ (b), and $0.22$ (c). The dashed lines are
the PMS isochrones from the models of Tognelli et al.  (2012), for ages
of 1, 2, 4, 8, 16, 32, and 64\,Myr, from right to left {(note that
where  multiple isochrones reach the MS and overlap, lines appear
solid instead of dashed)}. Also shown are  the representative
evolutionary tracks for masses of $0.7, 1, 1.5, 2, 3$  and 4\,\Msolar
(thin solid lines).} 
\label{fig5}
\end{figure}

\section{Physical properties of bona-fide PMS stars}

In order to select bona-fide PMS stars, after reddening correction in
all bands we repeated the procedure discussed in Section\,4 to identify
objects with $H\alpha$ excess emission but imposing stricter
constraints. In particular, we only considered stars with a
reddening-corrected $V-H\alpha$ excess at the $4\,\sigma$ level or
higher. Furthermore, to avoid contamination by stars with significant
chromospheric activity, we only considered objects with equivalent width
of the $H\alpha$ emission $W^E_{\rm eq}(H\alpha)> 20$\,\AA\, (for
details on deriving $W^E_{\rm eq}(H\alpha)$ from the photometry see e.g.
De Marchi et al. 2010, 2011a, 2013a). {We note that this high
threshold on $W^E_{\rm eq}(H\alpha)$ excludes from the sample all
weakly-accreting PMS stars present in the field. This is not a problem,
however, since our goal is to study the properties of the mass accretion
process in PMS stars. This requires a sample of bona-fide objects, but
completeness is not essential.}

This more stringent selection reduced the sample to 1\,035 stars with
effective temperature $T_{\rm eff} < 10\,000$\,K. They are shown as
filled dots in the Hertzsprung--Russell (H--R) diagrams in
Figure\,\ref{fig5}. Unfilled dots correspond to $\sim 400$ stars hotter
than 10\,000\,K that we will not consider in this work. These are
objects typically more massive than $\sim 2$\,\Msolar and might contain
Be stars that are evolving off the MS. Even though we could derive the
mass of these objects from their broad-band colours, without
spectroscopy their age would be rather uncertain because they may be
both pre- and post-MS stars. This, in turn, would make the comparison
more difficult with the lower-mass PMS stars of interest here. We will
not consider these objects further here and will address the properties
of these $H\alpha$ emitting stars in a future work (De Marchi \&
Panagia, in preparation).

{We note that amongst the objects with $H\alpha$ excess in
Figure\,\ref{fig5} there could be a very small fraction of evolved stars
with circumstellar envelopes, including planetary nebulae and post-AGB
stars. Objects in these phases, however, are short-lived, with typical
lifetimes of $\sim 10^5$\,yr (e.g. Vassiliadis \& Wood 1994; Habing
1997). This is about three orders of magnitude shorter than the
He-burning lifetime of stars in the RC phase ($\sim 10^8$\,yr, e.g.
Girardi \& Bertelli 1998; Girardi et al. 1998). With a total of about
180 RC stars in the field of our investigation (De Marchi \& Panagia
2014), one expects less than one of these post-AGB stars in our sample
of objects with $H\alpha$ excess. Their contribution is, therefore,
negligible and can be ignored. }

\subsection{Bolometric luminosity and effective temperature}

The value of the effective temperature $T_{\rm eff}$ was derived from
the dereddened $V-I$ colour, since in the range $4\,000 - 10\,000$\,K of
interest here this is an excellent index for temperature determinations 
(e.g. Bessell et al. 1998; von Braun et al. 1998; Romaniello 1998). The
models of Bessell et al. (1998) already discussed in Section\,3 were
used in the conversion. The bolometric luminosity ($L$) was derived from
the dereddened $V$ magnitude and the same models, having assumed a
distance modulus of $18.55$. The typical $\pm 1\,\sigma$ uncertainties
on the $T_{\rm eff}$ and $L$ are shown by the error bars on the
left-hand side of Figure\,\ref{fig5} (these uncertainties include the
effects of the reddening correction). Note that the uncertainty on
$T_{\rm eff}$ is on average larger for brighter stars, since they tend
to be hotter and the $V-I$ colour becomes increasingly less sensitive to
$T_{\rm eff}$ variations.

The three panels in Figure\,\ref{fig5} refer to the same objects, but in
each panel we have used a different reddening value for the stars
classified as older than 5\,Myr in the CMD of Figure\,\ref{fig3}. As
discussed in Section\,4, the values are $A_V=1.22$ (panel a), $0.72$
(b), and $0.22$ (c). The corresponding objects are marked with crosses.
For reference, we have traced in each panel the PMS isochrones (dashed
lines) from the models of Tognelli et al. (2012) for the assumed
$Z=0.007$ metallicity. Isochrones are shown for ages of 1, 2, 4, 8, 16,
32, and 64\,Myr, from right to left. The separation between isochrones is
typically larger than the uncertainties on $T_{\rm eff}$ and $L$ (see
error bars). Also shown are the representative evolutionary tracks for
masses of $0.7, 1, 1.5, 2, 3$ and 4\,\Msolar (solid curves) from the
same Tognelli et al. (2012) models.

The paucity of objects with $T_{\rm eff} \simeq 8\,000$\,K and $L \simeq
10\,L_\odot$ in panels b) and c) appears somewhat unphysical, and might
suggest that the reddening values adopted for the older PMS stars in
those cases are too low. A similar indication comes from the mass
distribution (see below). However, since one cannot a priori exclude
those lower $A_V$ values, we include the corresponding H--R diagrams
here for reference.

\subsection{Mass and age}

By means of a finer grid of models than the one shown in
Figure\,\ref{fig5}, we derived the individual masses and ages of the
bona-fide PMS stars. We followed the approach originally discussed by
Romaniello (1998) and more recently refined by De Marchi et al. (2011a,
2011b, 2013a). The method is similar to the one presented by Da Rio \&
Robberto (2012) and provides the probability distribution for each
individual star to have a given value of the mass and age, purely on the
basis of the measurement errors, and without making assumptions on the
properties of the population, such as the functional form of the initial
mass function.

{Briefly, the position of each star in the HR diagram is compared with
PMS evolutionary tracks in order to derive the most likely age and mass
for that object. The evolutionary tracks used are those of Tognelli et
al. (2011) mentioned above, interpolated to a finer mass grid with a
logarithmic step of $0.025$. 

We first define a grid in luminosity and temperature in the HR diagram,
with evenly spaced cells and logarithmic steps both in temperature and
luminosity. The sizes of the cells are chosen to be comparable to the
typical observational errors. Then, given the evolutionary model of a
star of a certain mass, we identify all the cells that that star crosses
throughout its evolution. For each cell, we compute the crossing time
(difference between the times of exit and entrance) and characteristic
age (average of exit and entrance times). We repeat this process for all
the tracks resulting from the interpolation mentioned above. 

The results are four matrices containing all the relevant information
associated with the evolutionary tracks crossing that cell, namely mass,
age, time of permanence in the cell, and multiplicity for that cell
(number of possible solutions). Given an observed star, the cell of the
grid to which that star belongs gives us the most likely mass and age of
the objects. If more than one solution is possible for that cell,
because more than one track crosses the cell or the same track crosses
it at different times, we assign an appropriate weight to each solution.
As zero-th order approximation, assuming a constant birthrate, the
weight is proportional to the time a given track spends in that cell.
Although this is typically sufficient for rapidly evolving PMS stars,
subsequent iterations are used to take into account a more realistic
star formation history. }

The largest source of uncertainty on the derived masses and ages comes
from the reddening correction. For younger PMS stars, where the
reddening correction is applied individually using the 
nearest-neighbours method, the typical uncertainty on $E(V-I)$ is about
$0.1$\,mag, as mentioned in Section\,4. This translates into a typical
uncertainty of $\sim 6$\,\% on the mass and of $\sim 18$\,\% on the age.
For older PMS objects, where all stars are assigned the same $A_V$
value,  we measured how varying $A_V$ over the full range from $0.22$ to
$1.22$ affects the individual masses and ages. Objects appear more
massive when $A_V$ grows and the typical mass variation introduced by a
spread of $0.5$\,mag in $A_V$ is about 8\,\%. The effect on the age
is about twice as large, corresponding to $\sim 19$\,\%.

\begin{figure}[t]
\centering
\resizebox{\hsize}{!}{\includegraphics{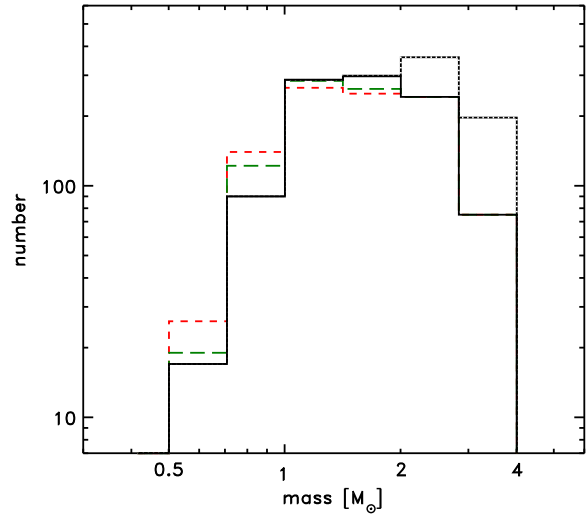}}
\caption{Histograms of the mass distribution for the 1\,035 bona-fide
PMS  stars. Different line types correspond to different $A_V$ choices
for the older PMS objects, respectively $A_V=1.22$ (solid line),
$A_V=0.72$ (long-dashed line), and $A_V=0.22$ (short-dashed line). For
completeness, the dotted line shows the mass distribution of the stars
with $H\alpha$ excess emission hotter than $10\,000$\,K that we do not
consider in this work.} 
\label{fig6}
\end{figure}

The histogram with the mass distribution for the 1\,035 PMS stars are
shown in Figure\,\ref{fig6}. The $\sqrt{2}$ binning has been
selected to match the evolutionary  tracks in Figure\,\ref{fig5} and is
several times wider than the typical uncertainty on the mass. Different
line types correspond to different $A_V$ choices for the older PMS
objects as mentioned above, respectively $A_V=1.22$ (solid line),
$A_V=0.72$ (long-dashed line), and $A_V=0.22$ (short-dashed line). The
histograms show a larger number of more massive stars when the value of
$A_V$ is higher, and correspondingly a smaller number of massive stars
when the value of $A_V$ is lower. However, the three mass distributions
are rather similar and statistically significant differences ($\sim
4\,\sigma$) are only seen in the mass range $0.7 - 1.0$\,\Msolar. 

The histograms show a relatively broad peak in the range $1 -
2$\,\Msolar. We underline that, while these measurements show the
relative mass distribution of PMS stars, they cannot set constraints on
the shape of the mass function, since we are only considering PMS stars
that had $H\alpha$ excess emission at the time of the observations and
we are also not accounting for the unavoidable photometric
incompleteness, {which is more severe at low masses.} 

\begin{figure}[t]
\centering
\resizebox{\hsize}{!}{\includegraphics[bb=20 0 484 395]{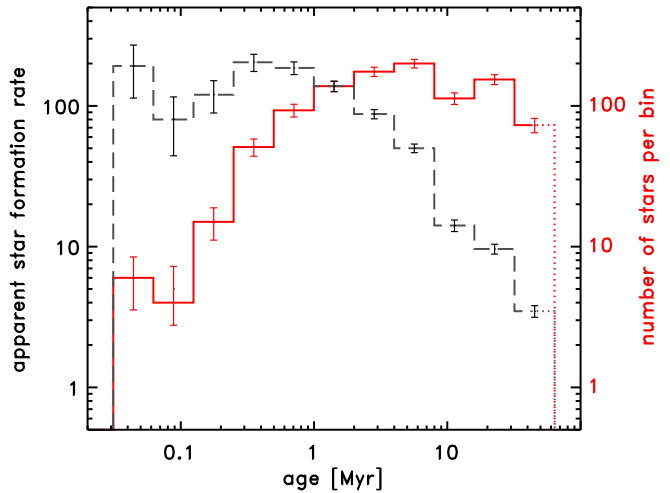}}
\caption{Histogram of the age distribution of PMS stars. A constant
logarithmic step is used (factor of 2). The solid line provides the number
of stars inside each age bin, while the dashed line shows the apparent
star formation rate in units of Myr$^{-1}$.} 
\label{fig7}
\end{figure}

The histograms in Figure\,\ref{fig7} show the age distribution for the
PMS stars, for an assumed $A_V=0.72$ for objects older than 5\,Myr. The
solid line gives the number of stars in each age bin, which we have
selected with a constant logarithmic step of a factor of 2 to match the
isochrones in Figure\,\ref{fig5}. The age bins are several times wider
than the uncertainty on the age. The dashed line gives the apparent
rate of star formation, obtained by dividing the number of stars in each
bin by the age extent of the bin. The figure can be directly compared
with Figure\,10 of De Marchi et al. (2011c) and it is consistent with
it: the apparent star formation rate in the past 2\,Myr is $ \sim
150$\,\Msolar\,Myr$^{-1}$ and it drops to $\sim 10$\,\Msolar\,Myr$^{-1}$
between 20 and 30\,Myr ago. 

These are necessarily lower limits to the star formation rate, because
we only consider here PMS stars with $H\alpha$ excess emission at the
$4\,\sigma$ level or more at the time of the observations. Moreover,
young stars show large variations in their $H\alpha$ emission over hours
or days (e.g. Fernandez et al. 1995; Smith et al, 1999; Alencar et al.
2001). Furthermore, an active state of accretion is expected to decrease
with age (e.g., De Marchi, Panagia \& Sabbi 2011; Fedele et al. 2010)
and this must be at least partly responsible for the rapid decline seen
at ages older than $\sim 2$\,Myr. In fact, if the fraction of PMS stars
with $H\alpha$ excess had an exponential decay with time (Fedele et al.
2010), episodes occurring more than 20\,Myr ago might have contributed
to a large portion of the stars in this region and the star formation
strength might have been considerably higher than that of the current
episode associated with R\,136. 

Adopting the reddening $A_V=0.22$ or $A_V=1.22$ for the older PMS stars
would have produced, respectively, a slightly steeper or shallower drop
of the apparent star formation rate for ages older than $\sim 10$\,Myr.
The differences, however, are not significant as they do not exceed
$1\,\sigma$. The reason is that in the CMD the isochrones are almost
parallel to one another for stars less massive than $\sim 1$\,\Msolar
and older than 10\, Myr, which are the objects of interest here.
Therefore, the reddening vector intersects the isochrones at a similar
angle, resulting in rather similar star formation rates. De Marchi et
al. (2011c), using the Galactic extinction law, found a somewhat gentler
drop of the apparent star formation rate beyond 10\,Myr (see their
Figure\,10). Qualitatively the drop is similar to the one we find here,
but Figure\,\ref{fig7} is quantitatively more accurate because it is
based on the extinction law appropriate for this region. 

We can thus confirm that in the central 30\,Dor regions star formation
is not limited to the most recent episode associated with the $\sim
3$\,Myr-old massive members of R\,136, but has been ongoing for at least
$\sim 30$\,Myr, proceeding from the outer layers towards the interior. An
accurate knowledge of the extinction law and of the value of $A_V$ are
crucial to obtain quantitative information and assign a mass and an age
to individual PMS stars. Nonetheless, some qualitative properties of
these populations, such as the star formation rate for older PMS stars
approaching the MS, do not appear to depend crucially on the precise
amount of the reddening. This is important in those regions where only
approximate information exists about extinction.     

\subsection{Accretion luminosity and mass accretion rate}

The $H\alpha$ luminosity $L(H\alpha)$ is obtained, after reddening
correction, from the difference between the $V-H\alpha$ colour of PMS
stars and the reference relation defined in Section\,3, for objects of
the same effective temperature. Also needed are the photometric
zero-point and absolute sensitivity of the instrumental setup (in this
case, the F656N filter of the WFC\,3), as well as the distance modulus.
The procedure is addressed in detail in De Marchi et al. (2010, 2011a,
2013a) and we refer the reader to those papers for further particulars. 

The main source of uncertainty on $L(H\alpha)$ is, once again, the
reddening correction. In the colour--colour diagram
(Figure\,\ref{fig2}), the reddening vector is substantially parallel to
the reference template (dashed line), so extinction has a practically
negligible effect on the $V-H\alpha$ colour excess and on the $H\alpha$
equivalent width. However, since $L(H\alpha)$ depends on the level of
the underlying continuum, any uncertainty on the latter will also affect
$L(H\alpha)$. The $0.1$\,mag uncertainty on $E(V-I)$ discussed earlier
corresponds to an uncertainty of $0.25$\,mag on the $H\alpha$ absolute
magnitude of the stars and hence a $\sim 22$\,\% uncertainty on the
$L(H\alpha)$ luminosity.

The median $L(H\alpha)$ is $\sim 3 \times 10^{-2}$\,L$_\odot$, with the
17 and 83 percentile levels in the distribution at $0.005$\,L$_\odot$
and $0.01$\,L$_\odot$, respectively. These values are in line with those
measured by De Marchi et al. (2011a) in the massive cluster NGC\,346 in
the Small Magellanic Cloud (SMC). On the other hand, they are about an
order of magnitude higher than those measured in the smaller SMC cluster
NGC\,602 (De Marchi et al. 2013a). The difference, however, is not
surprising because these observations probe stars of different masses. 

From $L(H\alpha)$ we can derive the accretion luminosity $L_{\rm acc}$,
which corresponds to the total energy released by the accretion process
that is assumed to result into ionising and heating of the circumstellar
gas (e.g., Hartmann et al. 1998). As discussed in De Marchi et al.
(2010, 2011a, 2013a), there is both observational and theoretical
evidence suggesting a linear dependence between $L_{\rm acc}$ and
$L(H\alpha)$, implying that a constant fraction of the gravitational
energy released in the accretion process goes into $H\alpha$ emission.
More recently, Ingleby et al. (2013) found a consistent slope of unity
not only for $L(H\alpha)$, but also for the luminosities in $H\beta$,
Ca~II~K, C~II], and Mg~II. 

De Marchi et al. (2010, 2011a, 2013a) found $\log L_{\rm acc} = (1.72 \pm
0.25) + \log L(H\alpha)$ from the analysis of a set of $L(H\alpha)$
measurements of a group of T Tauri stars in Taurus--Auriga compiled by
Dahm (2008) and based on a sample originally studied by Valenti et al.
(1993). We will use this relationship to derive $L_{\rm acc}$ from our
PMS objects from the observed $L(H\alpha)$ values. It is interesting to
note that this proportionality factor between $L_{\rm acc}$ and
$L(H\alpha)$ corresponds to what one would expect for the ionising
radiation of the atmosphere of a $\sim 40\,000 - 50\,000$\,K star (see
Panagia 1973). While this finding does not per se identify the specific
location where the energy is produced (e.g. an accretion shock, a
stellar transition region, a shock in the stellar outflow, etc.), it
suggests the existence of a heated region close to the photosphere (for
a recent review of the emission mechanisms see Ardila et al. 2013).

The statistical uncertainty on $L_{\rm acc}$ is dominated by the quoted
22\,\% uncertainty on $L(H\alpha)$ associated with the correction for
reddening. There is a larger systematic uncertainty, namely a factor of
$\sim 1.8$, due to the poorly known ratio $L_{\rm acc} / L(H\alpha)$
{(see Dahm 2008; De Marchi et al. 2011a)}, but since this ratio is
the same for all stars, this uncertainty will not hamper the comparison
between different objects.

Once $L_{\rm acc}$ is known, we can derive the mass accretion rate $\dot
M_{\rm acc}$. The free-fall equation links the luminosity released by the
impact of the accretion flow with the rate of mass accretion, according
to the relationship

\begin{equation}
L_{\rm acc} \simeq \frac{G\,M_*\,\dot M_{\rm acc}}{R_*} \left(1 -
\frac{R_*}{R_{\rm in}}\right)
\label{eq1}
\end{equation}

\noindent 
where $G$ is the gravitational constant, $M_*$  the star's mass
determined above and $R_*$ its photospheric radius, obtained from its
luminosity and effective temperature. $R_{\rm in}$ is the inner radius
of the accretion disc and its value is rather uncertain. We follow
Gullbring et al. (1998) and adopt $R_{\rm in} = 5\,R_*$ for all PMS
objects. 

\begin{figure}[t]
\centering
\resizebox{\hsize}{!}{\includegraphics{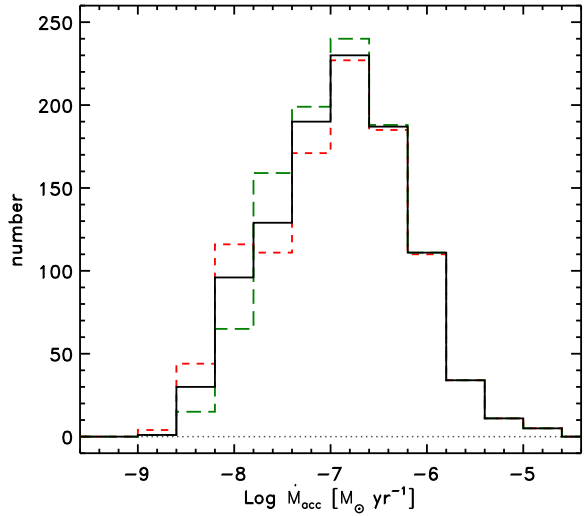}}
\caption{Histograms of the mass accretion rate values of the bona-fide
PMS stars. As before, different line types correspond to different $A_V$
choices for the older PMS objects, respectively $A_V=1.22$ (solid line),
$A_V=0.72$ (long-dashed line), and $A_V=0.22$ (short-dashed line).}
\label{fig8}
\end{figure}

Histograms of the measured values of the mass accretion rate are shown
in Figure\,\ref{fig8}, and the median value in our sample is $\sim 8
\times 10^{-8}$\,\Msolar\,yr$^{-1}$. The solid line refers to $A_V=0.72$
for the  PMS stars older than $\sim 5$\,Myr, while the short-dashed and
long-dashed lines correspond, respectively, to $A_V=0.22$ and
$A_V=1.22$.  

The main source of statistical uncertainty on $\dot M_{\rm acc}$ is,
again, the  correction for reddening, which introduces an uncertainty of
$\sim 22$\,\% on $L_{\rm acc}$ as mentioned above. Also the ratio
$R_*/M_*$ is sensitive to uncertainties on the reddening correction, but
to a smaller extent. De Marchi et al. (2010) have shown that, in the
H--R plane, the reddening vector is almost parallel to the lines of
constant radius (see their Figures\,10 and 11), and therefore extinction
affects only marginally the estimate of the stellar radius. However,
since at low temperatures the reddening vector crosses the evolutionary
tracks, any unaccounted extinction would lead to an underestimate of the
stellar mass, thus resulting in a systematically higher value of the
$R_*/M_*$ ratio. On the other hand, this effect is relatively small, and
a typical uncertainty of $0.1$\,mag  in $E(V-I)$, as relevant in our
case, corresponds to an uncertainty of $\sim 6$\,\% on the $R_*/M_*$
ratio. In summary, the combined statistical uncertainty on $\dot M_{\rm
acc}$ amounts to $\sim 23$\,\%. When also the systematic uncertainties
due to the conversion between $L(H\alpha)$ and $L_{\rm acc}$ are
considered, the final combined uncertainty on $\dot M_{\rm acc}$ amounts
to less than a factor of 2. This is important, because it allows us to
make a solid  comparison  between the $\dot M_{\rm acc}$ values for
objects of different masses and ages across our sample.

\section{Mass accretion rate and stellar parameters}

The broad distribution of $\dot M_{\rm acc}$ values seen in
Figure\,\ref{fig8} reflects the wide range of masses $m$ and ages $t$
covered by this sample. We will now explore how $\dot M_{\rm acc}$
depends on both $m$ and $t$ for these stars, and later will compare
these results with other massive star forming regions in the MCs and MW.
The large size of our samples of PMS stars will allow us to derive
statistically significant conclusions. 

\begin{figure}[t]
\centering
\resizebox{\hsize}{!}{\includegraphics{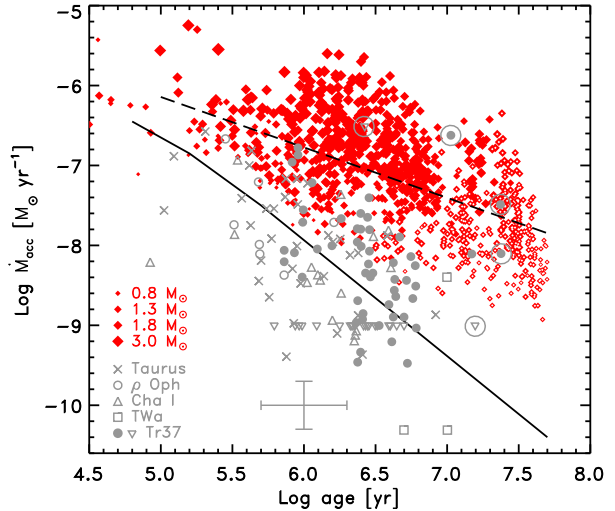}}
\caption{Mass accretion rates of bona-fide PMS stars, as a function of
stellar age. Diamonds refer to the objects in 30\,Dor and their size is 
proportional to the stellar mass (see legend). In this graph, we have
adopted $A_V=0.72$ for all PMS stars older than 5\,Myr (unfilled
diamonds). The dashed line represents the best fit to the 30\,Dor data. 
Other symbols correspond to a sample of Galactic T-Tauri stars from the
literature, as per the legend (the large circles  around some of the
symbols indicate dwarfs of spectral type G in the Trumpler 37 sample).
The solid line shows the relationship between $\dot M_{\rm acc}$ and $t$
predicted by models of the evolution of a viscous disc (Hartmann et al.
1998).} 
\label{fig9}
\end{figure}

We begin by combining information on the three parameters in one graph,
showing in Figure\,\ref{fig9} the values of $\dot M_{\rm acc}$ as a
function of $t$. The 30\,Dor PMS stars are marked by diamonds and the 
size of the symbols is proportional to the mass (see legend). As a
representative value of the extinction for PMS stars older than 5\,Myr
(unfilled diamonds) we have assumed $A_V=0.72$. Overall, the mass
accretion rate of the 30\,Dor objects appears to decline with age as
$\dot M_{\rm acc} \propto t^{-0.63}$, as shown by the best fit to the
data (dashed line). The other symbols, as per the legend, correspond to
a sample of Galactic T-Tauri stars studied by Sicilia-Aguilar et al.
(2006, 2010), and the typical uncertainty on their values is shown by
the large cross at the bottom. These objects suggest a steeper drop of
$\dot M_{\rm acc}$ with $t$, consistent with the models of viscous disc
evolution by Hartmann et al. (1998) shown by the solid line. 

The observed slope ($\alpha = -0.63 \pm 0.02$) is in good agreement with
those measured in other MCs environments, namely NGC\,346
($\alpha=-0.55$; De Marchi et al. 2011a) and NGC\,602 ($\alpha=-0.7$; De
Marchi et al. 2013a) in the SMC. However, it is considerably shallower
than the $\alpha \simeq -1.5$ slope predicted by the models of Hartmann
et al. (1998; see also Calvet et al. 2000; Muzerolle et al. 2000). A
steeper decline, consistent with these models, has also been observed by
Barentsen et al. (2011) and Beccari et al. (2015) for low-mass stars ($<
0.6$\,\Msolar) in Galactic clusters, respectively in Tr\,37 in Cepheus
and Tr\,14 in the Carina Nebula. 

Therefore, there might be differences between the MCs and the Galaxy
environments affecting how the mass accretion rate scales with age.
However, to properly explore this possibility, we cannot simply rely on
graphs of the type of Figure\,\ref{fig9}, since the mass spread present
at all ages in that figure contributes to the apparent scatter. In
fact, $\dot M_{\rm acc}$ is also expected to depend on $m$. In the
low-mass regime, this dependence could be as steep as $m^2$ (Muzerolle
et al. 2003, 2005; Natta et al. 2004, 2006; Calvet et al. 2004), or even
steeper in the very low-mass domain (Fang et al. 2009). On the other
hand, those conclusions are based on rather limited samples of a few
dozens PMS stars each, often with a significant age spread. In the
Appendix, we address how an age spread affects the inferred mass dependence
of $\dot M_{\rm acc}$ and results in erroneously steep indices.

\begin{figure}[t]
\centering
\resizebox{1.1\hsize}{!}{\includegraphics{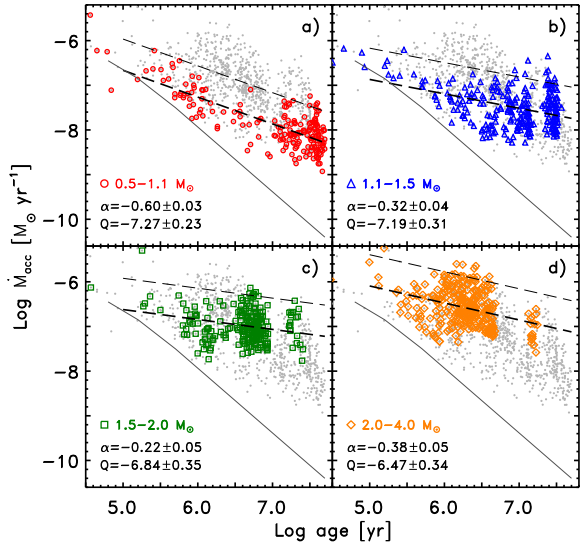}}
\caption{Same as Figure\,\ref{fig9} but for stars in different mass
groups,  as indicated in each panel. The slope $\alpha$ and intercept
$Q$ (at 1\,Myr)  of the best fits (thick dashed lines) are given in each
panel. The values of $\alpha$ are considerably shallower than the slope
predicted by the models of Hartmann et al. (1998; solid line). {The
thin dashed lines represent the lines of best fit shifted vertically by
$0.7$\,dex and define good envelopes to the observed distributions.}}
\label{fig10}
\end{figure}

Figure\,\ref{fig10} shows the values of $\dot M_{\rm acc}$ in 30\,Dor as
a function of age $t$ for four roughly equally populated mass groups,
namely, $0.5-1.1$\,\Msolar, $1.1-1.6$\,\Msolar, $1.6 - 2.0$\,\Msolar,
and $2.0 - 4.0$\,\Msolar. Also given in each panel are the slope
$\alpha$ and intercept $Q$ (at 1\,Myr) of the best linear fit to the
data (dashed lines), according to the relationship $\log \dot M_{\rm
acc} = \alpha \times \log(t) + Q$ with the age in Myr. The values of the
slope $\alpha$ range from $\sim -0.2$ to $\sim -0.6$ and are in relative
agreement with one another, (mean $-0.39$ and standard deviation
$0.13$), but are all considerably shallower than the $\sim - 1.5$ slope
predicted by the models of Hartmann et al. (1998; solid line in the
figure). {We note that our conservative selection of only the
bona-fide PMS stars with $W^E_{\rm eq}(H\alpha)> 20$\,\AA\, does not
appear to affect the measured slopes. As already discussed in detail in
Paper II for the case of NGC\,346, the upper envelopes of the
distribution of all four mass groups appear to be fully consistent with
the slope of the best fit in each panel. To highlight this, the thin
long-dashed lines shown in the figure represent the lines of best fit
shifted vertically by $0.7$\,dex. Therefore, even if our conservative
detection limits do not allow us to identify objects with low
$L(H\alpha)$, no selection effects are to be expected for stars with
high $H\alpha$ excess. Since the distribution of these objects is
consistent with the slope of the best fit, the latter is not skewed by 
having ignored objects with the smallest mass accretion rates, and
therefore it must be correct.} 

The dependence of the mass accretion rate on the stellar mass is
described in a qualitative way by the systematic growth of $Q$ with
increasing mass. However, the values of $Q$ listed in
Figure\,\ref{fig10} cannot be directly compared to one another to find
the formal dependence of $\dot M_{\rm acc}$ on $m$, because they refer
to different slopes $\alpha$. In this case, assuming that all stars in
our sample formed under the same conditions, a proper and quantitative
analysis is only possible through a multivariate least-squares fit that
takes into account the simultaneous dependence of $\dot M_{\rm acc}$ on
both $m$ and $t$. This approach is allowed by the large size of our PMS
stars sample. Adopting a simple relationship of the type 

\begin{equation}
\dot M_{\rm acc} \propto t^a \times m^b,
\label{eq2}
\end{equation}

\noindent 
De Marchi et al. (2011a, 2013a) found $a=-0.59 \pm 0.02$ and $b=0.82 \pm
0.09$ for NGC\,346 and $a=-0.72 \pm 0.02$ and $b=0.94 \pm 0.14$ for
NGC\,602, both in the SMC. These values apply to objects in the mass
range $0.5 - 1.5$\,\Msolar and younger than $\sim 16$\,Myr. If we
consider the $\sim 300$ objects with these characteristics in our
30\,Dor sample, we find $a=-0.55 \pm 0.03$ and $b=1.25 \pm 0.19$. De
Marchi et al. (2013a) already pointed out that approximate values of
$a=-0.6$ and $b=1$ result in fits with small residuals for both NGC\,346
and NGC\,602 and this is confirmed also for 30\,Dor.

\section{Mass accretion rate and environment}

Comparing the mass accretion properties of the stars in 30\,Dor with
those in NGC\,346, NGC\,602, and other star-forming regions in the MCs
and in the Galaxy can provide quantitative information on the effects of
the environment on the final stages of star formation. To account for
differences in the environment, it is convenient to rewrite the
power-law dependence on mass and age in logarithmic form, namely

\begin{equation}
\log \dot M_{\rm acc} = a \times \log \frac{t}{{\rm Myr}} + b \times \log
\frac{m}{M_\odot} + c. 
\label{eq3}
\end{equation}

If the mass accretion rate were only a function of $t$ and $m$, the term
$c$ would be a constant, representing in practice an ``effective mass
accretion rate'', namely the mass accretion rate of a 1\,\Msolar PMS
star with an age of 1\,Myr. In practice, however, the best fitting value
of the $c$ term in each region will reflect environmental effects, which
are not considered explicitly in Equation\,\ref{eq3}, such as for
instance the metallicity, the mean gas density, or the strength of the
local magnetic field. The effects of metallicity are particularly
interesting, since in general a lower metallicity implies a lower
opacity, temperature, and viscosity for the disc, and thus a longer
viscous time (e.g. Durisen et al. 2007).

\begin{deluxetable*}{llrcccccccc}
\tablecolumns{11}
\tabletypesize{\footnotesize}
\tablecaption{Parameters of the selected regions.\label{tab1}} 
\tablehead{\colhead{Region} & \colhead{Reference} & \colhead{Stars} & 
\colhead{$Z$} & \colhead{$a$} & \colhead{$b$} & \colhead{$c$} &
\colhead{$a^\prime$} & \colhead{$b^\prime$} & \colhead{$c^\prime$} & 
\colhead{$d^\prime$} \\[0.05cm]
\multicolumn{1}{c}{(1)} & \multicolumn{1}{c}{(2)} & (3) & (4) & (5) & (6) & 
(7) & (8) & (9) & (10) & (11) }

\startdata
NGC\,346 & De Marchi et al. (2011a) &  375 & $2$ & 
$-0.59\pm0.02$ & $0.78\pm0.08$ & $-3.41\pm0.02$ & 
$-0.61\pm0.02$ & $0.74\pm0.08$ & $-0.29\pm0.02$ & $-7.22 \pm 0.02$ \\ 
NGC\,602 & De Marchi et al. (2013a)  &  105 & $4 - 15$ &
$-0.59\pm0.02$ & $0.78\pm0.08$ & $-3.78\pm0.04$ & 
$-0.61\pm0.02$ & $0.74\pm0.08$ & $-0.29\pm0.02$ &$-7.43 \pm 0.04$ \\ 
30\,Dor  & this work                &  291 & $7$ &
$-0.59\pm0.02$ & $0.78\pm0.08$ & $-3.67\pm0.02$ & 
$-0.61\pm0.02$ & $0.74\pm0.08$ & $-0.29\pm0.02$ & $-7.32 \pm 0.02$ \\
SN\,1987A& De Marchi et al. (2010)  &  ~55 & $7$ &
$-0.59\pm0.02$ & $0.78\pm0.08$ & $-3.47\pm0.02$ & 
$-0.61\pm0.02$ & $0.74\pm0.08$ & $-0.29\pm0.02$ & $-7.11 \pm 0.02$ \\
Tr\,14   & Beccari et al. (2015)    &  237 & $19$ &
$-0.59\pm0.02$ & $0.78\pm0.08$ & $-3.72\pm0.02$ & 
$-0.61\pm0.02$ & $0.74\pm0.08$ & $-0.29\pm0.02$ & $-7.26 \pm 0.02$ \\
NGC\,3603& Beccari et al. (2010)    &  244 & $19$ &
$-0.59\pm0.02$ & $0.78\pm0.08$ & $-3.65\pm0.02$ & 
$-0.61\pm0.02$ & $0.74\pm0.08$ & $-0.29\pm0.02$ & $-7.19 \pm 0.02$ \\
\enddata
\tablecomments{Table columns are as follows: (1) names of the regions;
(2) works from which the PMS catalogues are taken; (3) number of PMS stars
in each region with mass in the range $0.5 - 1.5$\,\Msolar and younger
than $\sim 16$\,Myr; (4) metallicity in units of $10^{-3}$; (5 -- 7)
best fitting parameter values as per Equation\,\ref{eq3}; (8 -- 11) best
fitting parameter values as per Equation\,\ref{eq4}.}   
\end{deluxetable*}

Besides 30\,Dor, NGC\,346 and NGC\,602, we consider PMS stars in the
SN\,1987A field (De Marchi et al. 2010), in Tr 14 (Beccari et al. 2015)
{and from an ongoing study in NGC\,3603 (De Marchi, Beccari \&
Panagia, in prep.; see also Beccari et al. 2010).} PMS objects in these
regions have been identified and characterised in a homogeneous way from
their $H\alpha$ excess emission, so they represent a particularly
uniform sample. Following the same approach, Spezzi et al. (2012)
studied three more regions in the LMC located between 30\,Dor and the
SN\,1987A field. However, we do not include them in the present study
since the extinction law adopted in that work is now superseded
(Ma{\'\i}z Apell{\'a}niz et al. 2014; De Marchi \& Panagia 2014; De
Marchi et al. 2016), likely resulting in  artificially inflated values
of the mass accretion rate. 

Details of the {6 regions} considered here are listed in
Table\,\ref{tab1}. Besides the names of the regions (column 1), we
indicate the works from which the PMS catalogues are taken (column 2),
the number of stars in each sample (column 3), the metallicity (column
4), and the values of the best-fitting parameters resulting from
Equations\,\ref{eq3} and \ref{eq4} (columns 5--11). 

Concerning the metallicity, we consider a wide range of values for
NGC\,602. This cluster and associated N\,90 nebula are in the SMC Wing,
for which Lee et al. (2005) derived a chemical composition similar to
that of other SMC stars, based on the analysis of four B-type
supergiants. Therefore, in the literature (see, e.g., Carlson et al.
2007; Nigra et al. 2008; Cignoni et al. 2009; Gouliermis et al. 2012;
Oskinova et al. 2013; De Marchi et al. 2013a) it is customary to adopt
for NGC\,602 the same metallicity as for the SMC, namely $Z=0.004$.
However, there is no proper direct measurement of the metallicity of
NGC\,602 itself. From optical spectroscopy of Sk\,183, the brightest
ionising star in NGC\,602, Evans et al. (2012) derived a N abundance
close to the solar value, namely $12+\log([N/H])=7.73 (\pm 20\,\%)$
compared to $7.83$ for the sun (Asplund et al. 2009), and as such
significantly richer than the typical values found in H\,II regions in
the SMC ($\sim 6.55$; Russell \& Dopita 1990). Thus, adopting $Z_\odot =
0.019$ (see e.g. Asplund et al. 2009; Lodders 2010; Caffau et al. 2010,
2011), the Sk\,183 metallicity value is $\sim 0.015$. Although the N
enrichment in Sk\,183 could in part result from CNO-processed gas
dredged up to the stellar surface, as suggested for Sher\,25 or the
progenitor to SN\,1987A (e.g., Smartt et al. 2002), it is quite possible
that the metallicity of the NGC\,602 region is considerably higher than
the SMC average. Thus, we will consider for NGC\,602 a range of values
between $Z=0.004$ and  $Z=0.015$.

We restrict our investigation to active PMS stars (i.e. those currently
undergoing mass accretion) with masses in the range $0.5 - 1.5$\,\Msolar
and younger than $\sim 16$\,Myr. The combined sample includes 1\,307
objects (see Table\,\ref{tab1}).

{The parameter values providing the best simultaneous fit as per
Equation\,\ref{eq3} are $a=-0.59 \pm 0.02$ and $b=0.78 \pm 0.08$
(columns 5 and 6 in Table\,\ref{tab1}), with a reduced $\chi^2=1.33$}.
Using these parameter values, we derive for each region the value of $c$
listed in Table\,\ref{tab1} (column 7). As mentioned above, $c$ can be
used to explore the effects of the environment on the mass accretion
rate. Available data were already indicating that the physical
conditions of the environment have an effect on the extent and duration
of the star formation process, since the mass accretion rate for stars
of the same mass and age appears to be systematically higher in the MCs
than in the Galaxy, (De Marchi et al. 2011a, 2013a). It is, thus, very
interesting to investigate whether there is an explicit dependence of
$\dot M_{\rm acc}$ on metallicity.

\begin{figure}[t]
\centering
\resizebox{\hsize}{!}{\includegraphics{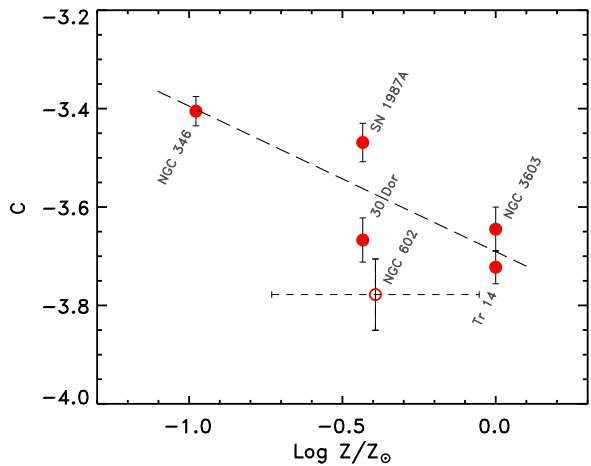}}
\caption{Effective mass accretion rate $c$ shown as a function of the
metallicity of the regions in our sample (see Table\,\ref{tab1}). Error
bars correspond to $\pm 2\,\sigma$.} 
\label{fig11}
\end{figure}

Indeed, the run of $c$ as a function of metallicity shown in
Figure\,\ref{fig11} reveals a correlation between these quantities (the
metallicity is shown in solar units, having adopted $Z_\odot=0.019$).
{To better quantify the level of correlation, we compute its
Pearson's correlation coefficient. This parameter does not allow us to
account for uncertainties on the data points. Since an accurate
metallicity for NGC\,602 is not yet known and a very wide range of
values is possible, we will ignore this region when calculating the
coefficient. In this case, the correlation is significant to better than
the 10\,\% level ($p=0.082$). If the metallicity of NGC\,602 were close
to $Z=0.015$, as the only available spectroscopic abundance measurements
suggest (Evans et al. 2012), the correlation would be statistically
significant at the 5\,\% level ($p=0.047$).  The formal best linear fit,
obtained ignoring NGC\,602, is $c = (-3.69 \pm 0.02) - (0.30 \pm 0.04)
\times \log Z/Z_\odot $ (dashed line). }

In light of the metallicity trend revealed by Figure\,\ref{fig11}, we
can rewrite Equation\,\ref{eq3} in a format that explicitly includes
the metallicity dependence, namely

\vspace*{-0.5cm}
\begin{equation}
\log \dot M_{\rm acc} = a^\prime \times \log \frac{t}{{\rm Myr}} +
b^\prime \times \log \frac{m}{M_\odot} + c^\prime \times \log \frac{Z}{Z_\odot}
 + d^\prime. 
\label{eq4}
\end{equation} 

We have fitted this relationship to all selected 1\,307 stars in the
regions listed in Table\,\ref{tab1}. As for the values of the
parameters  that simultaneously best fit all regions, the dependence on
mass and age remain practically unchanged {($a^\prime=-0.61 \pm 0.02$,
$b^\prime=0.74 \pm 0.08$), and $c^\prime=-0.29 \pm 0.02$, as indicated
in Table\,\ref{tab1} (columns 8, 9, 10).} The values of the parameter
$d^\prime$ are also listed in Table\,\ref{tab1} (column 11). The reduced
$\chi^2$ is $1.12$, suggesting that indeed the metallicity could play an
important role in determining the mass accretion rate of a star. The
observations do not reveal the exact mechanism, but one possible
explanation for the role of metallicity on the rate of mass accretion,
and hence on the intensity and duration of the accretion process, is
that the lower radiation pressure exercised by the forming star on
low-metallicity disc material is less efficient at dispersing the disc.
Thus, the accretion process may remain active for a longer time in
lower-metallicity environments. 

However, metallicity alone might not be sufficient to explain all the 
differences that we see in the mass accretion rates in our sample, 
particularly if the true metallicity of NGC\,602 is close to the lower
end of the range considered in Table\,\ref{tab1} ($Z\simeq 0.004$). In that
case, the difference between the expected and measured values of $c$
exceeds a $5\,\sigma$ deviation and suggests that there might be
additional environmental effects causing a characteristically lower mass
accretion rate in this region. 

One possibility is lower gas density. NGC\,602 is indeed a relatively
low-mass star-forming region located outside the main body of the SMC,
as mentioned above, in the so-called SMC Wing, a region characterised by
a moderate neutral hydrogen (HI) column density. Nigra et al. (2008)
discussed the environment of NGC\,602, pointing out that the large
number of galaxies visible both outside and within the nebula indicate
that the region is generally transparent, i.e. it lacks large-scale dust
and molecular concentrations. On the basis of photometry,
high-resolution echelle spectroscopy, and 21\,cm HI spectrum survey
data, they concluded that NGC\,602 formed at the periphery of a
low-density HI cloud, under conditions that are marginal for star
formation. 

The other regions listed in Table\,\ref{tab1} formed in typically denser
environments, as witnessed for instance by the higher intra-cluster
extinction values (for NGC\,346 see e.g. Sabbi et al. 2007, Hennekemper
et al. 2008, De Marchi et al. 2011a; for NGC\,602 see e.g. Carlson et
al. 2007,  Schmalzl et al. 2008, De Marchi et al. 2013a; for the
SN\,1987A field see e.g. Romaniello 1998, Panagia et al. 2000,
Romaniello et al. 2002; for 30\,Dor see e.g. Ma{\'\i}z Apell{\'a}niz et
al. 2014, De Marchi \& Panagia 2014, De Marchi et al. 2016; for Tr\,14
see e.g. Sana et al. 2010, Hur et al. 2012, Beccari et al. 2015).
Therefore, it is possible that the average lower mass density in
NGC\,602 resulted in less massive and more anemic circumstellar discs,
in turn implying more modest accretion rates. In order to address this
possibility in more detail, we plan to derive more accurate metallicity
information for NGC\,602 from spectroscopic measurements of the O/H
ratio.

\section{Summary and conclusions}

We have studied the properties of the young stellar populations in a
field of $\sim 2\farcm7 \times 2\farcm7$ around the centre of 30\,Dor
that was observed in 2009 October with the WFC\,3 on board the HST and
originally studied by De Marchi et al. (2011c). In order to identify PMS
stars still actively accreting matter from their circumstellar discs, we
have employed the photometric detection method described in detail
Papers\,I and II of this series (De Marchi et al. 2010; 2011a). Without
requiring spectroscopic observations, this method allows us to combine
broadband $V$ and $I$ photometry with narrow-band $H\alpha$ imaging in
order to identify stars with $H\alpha$ excess emission, to measure their
accretion luminosity $L_{\rm acc}$, and to derive their mass accretion
rate $\dot M_{\rm acc}$. The results of our analysis of 30\,Dor can be
summarised as follows.

\begin{enumerate}

\item   
From the original photometric catalogue of De Marchi et al. (2011c),
updated according to the most recent WFC\,3 photometric zero points, we
extracted about 18\,800 sources detected simultaneously in the $V$
and $I$ bands with photometric uncertainties of less than $0.1$\,mag in
both bands. About $2\,175$ of these stars have a $V-H\alpha$ colour
exceeding that of normal stars by at least $0.1$\,mag or by three times
the photometric uncertainty on the $V-H\alpha$ colour (i.e.
$3\,\sigma$), whichever is larger, and their combined photometric
uncertainty in all three bands does not exceed $0.15$\,mag. This is our
initial sample of PMS candidates. Their locations across the CMD
indicate a spread of ages exceeding 10\,Myr. 

\item  
We corrected for extinction the photometry of all young stars in
the field (i.e. PMS and UMS objects) using the most recently derived
extinction law specific for 30\,Dor (Ma{\'\i}z Apell{\'a}niz et al.
2014; De Marchi \& Panagia 2014; De Marchi et al. 2016). This field is
known to contain significant differential extinction ($0 \la A_V \la 3$)
and to display an extinction law considerably different from the
standard Galactic one.

\item
Younger PMS stars ($\la 5$\,Myr) and UMS stars share a very similar
spatial distribution, so we used the closest reference UMS stars to
derive an individual reddening correction for each younger PMS object.
The typical uncertainty on this reddening correction is $0.1$\,mag on
$E(V-I)$ or $\sim 0.3$\,mag on $A_V$. Older PMS stars ($\ga 5$\,Myr)
have a rather uniform spatial distribution and a more limited intrinsic
reddening spread ($0 \la A_V \la 1$), so we applied one and the same
correction to all stars. We adopted the average value of $A_V=0.5$ and
discussed the consequences of using $A_V=0$ or $A_V=1$ instead. For all
PMS stars, we also account for the additional $A_V=0.22$ component due
to the intervening Galactic absorption along the line of sight.  

\item 
We find that the older PMS candidates are located in front of the
R\,136 cluster and are separated from it by a conspicuous amount of
absorbing material. Besides a more  uniform spatial distribution, the
maximum total extinction towards the older PMS objects ($A_V=1.22$) is
considerably lower than the extinction towards the younger PMS stars (17
percentile $A_V=1.63$). The observations indicate that star formation
has proceeded from the periphery into the interior of the region.

\item  
The final catalogue of bona-fide PMS stars contains 1\,035 objects. We
determined their effective temperatures, luminosities, and radii by
comparing their photometry to the stellar atmospheric models of Bessell et
al. (1998). The masses and ages were derived by comparing their
luminosities and effective temperatures with the PMS evolutionary models
of the Pisa group (Tognelli et al. 2012) for an adopted metallicity
$Z=0.007$. The masses are in the range $0.5 - 4$\,\Msolar with a median
of $1.6$\,\Msolar. The median age is $4.2$\,Myr, but the sample spans
the range $\sim 0.1 - 50$\,Myr, confirming that star formation was
active in this field well before the recent episode giving birth to the
R\,136 cluster $\sim 2 - 3$\,Myr ago.  

\item 
We set a lower limit to the star formation rate for the current ($\la
2$\,Myr) burst of $\sim 150$\,\Msolar\,Myr$^{-1}$, while at $\sim
20$\,Myr the apparent rate drops to $\sim 10$\,\Msolar\,Myr$^{-1}$.
These are  both necessarily lower limits, since they include only stars
in the range  $0.5 - 4.0$\,\Msolar\, that were undergoing active mass
accretion at the time of the observations. Furthermore, if the fraction
of PMS stars with $H\alpha$ excess decays exponentially with time, the
episode or episodes older than $\sim 20$\,Myr ago might have been
considerably stronger than the current one.

\item
From the measured $H\alpha$ luminosity of these PMS stars we derived the
accretion luminosities $L_{\rm acc}$ and, through the other stellar 
physical parameters, also their mass accretion rates $\dot M_{\rm
acc}$. The sample includes objects with $1.5 \times
10^{-9}$\,\Msolar\,yr$^{-1} \la \dot M_{\rm acc} \la 1.5 \times
10^{-5}$\,\Msolar\,yr$^{-1}$. Our stringent selection of stars with
$H\alpha$ excess emission above the $4\,\sigma$ level limits the
identification of objects with lower $\dot M_{\rm acc}$ values that
are expected to exist in this field. 

\item
The dominant source of statistical uncertainty on the stellar physical
parameters is the reddening correction. The typical $0.1$\,mag
uncertainty on $E(V-I)$ translates to an uncertainty of $\sim 6\,\%$ on
the mass, $\sim 18\,\%$ on the age, $\sim 22\,\%$ on $L(H\alpha)$ and 
$L_{\rm acc}$, and $\sim 6\,\%$ on the stellar radius-to-mass ratio. The
combined statistical uncertainty on $\dot M_{\rm acc}$ amounts to $\sim
23\,\%$. When also the systematic uncertainties on the conversion
between $L(H\alpha)$ and $L_{\rm acc}$ are considered, the final
combined uncertainty on $\dot M_{\rm acc}$ amounts to  less than a
factor of 2. This permits a solid comparison between the $\dot M_{\rm
acc}$ values for objects of different masses and ages across our sample.

\item
We studied the relationship between the mass accretion rate and the mass
and age of the PMS stars in our sample. We show that attempts to derive
correlations by simply fitting separately the observed dependence of
$\dot  M_{\rm acc}$ on mass and age may introduce large biases and
result in indices that do not have general validity, because the
combined effects of age on both $\dot M_{\rm acc}$ and the range of
masses in the sample are ignored. Since these effects are intertwined, a
proper multivariate fit is needed. A linear regression fit of the type
$\log \dot M_{\rm acc} \simeq a \times \log t + b \times \log m + c$,
where $t$ is the age of the star in Myr and $m$ its mass in \Msolar,
gives $a=-0.55 \pm 0.03$ and $b=1.25 \pm 0.19$ for 30\,Dor. These values
are compatible with those found for NGC\,346 and NGC\,602 in the SMC.

\item
We applied the same multivariate regression fit to a uniform sample of 
1\,307 PMS stars obtained by selecting all objects with masses of 
$0.5-1.5$\,\Msolar and younger than $\sim 16$\,Myr contained in six
different star-forming regions, namely 30\,Dor and the SN\,1987A field
in the LMC, NGC\,346 and NGC\,602 in the SMC, and Tr\,14 {and
NGC\,3603} in the Galaxy. The values of the parameters $a$ and $b$ that
simultaneously best fit {all regions are $a=-0.59\pm0.02$ and
$b=0.78\pm0.08$, with $\chi^2=1.33$.} The values of $a$ and $b$ over the
entire sample converge to values that, within the uncertainties, are
quite close to those of the individual regions.

\item
The residuals are systematically different between the six regions
and they reveal a strong correlation with metallicity, of the type
$c = (-3.69 \pm 0.02) - (0.30 \pm 0.04) \times \log Z/Z_\odot$. If the
metallicity dependence is explicitly included in a multivariate fit of
the type $\log \dot M_{\rm acc} \simeq a \times \log t + b \times \log m
+ c^\prime \times \log Z + d^\prime$, {the best fitting parameter
values remain practically unchanged ($a^\prime=-0.61 \pm 0.02$,
$b^\prime=0.74 \pm 0.08$, $c^\prime=-0.29 \pm 0.02$) and the
corresponding reduced $\chi^2$ drops to $1.12$.} This finding confirms
the validity of the observed dependence of $\dot M_{\rm acc}$ on
metallicity.  

\item
A possible interpretation of the observed trend with metallicity is that
when the metallicity is higher there are more grains in the disc and
therefore radiation pressure limits the accretion process, both in its
rate and duration. 

\end{enumerate}

In conclusion, the analysis of the data shown in Figure\,\ref{fig11} and
Table\,\ref{tab1} reveals that the general properties of the
environment, such as the metallicity and possibly the gas density, can
affect appreciably the accretion process on a global scale. Obviously,
the evolution and mass accretion history of individual PMS stars can
still vary widely within the same region due to effects such as the
proximity of massive ionising stars (e.g. De Marchi et al. 2010; Zeidler
et al. 2016) or tidal interactions (e.g., Olczak, Pfalzner \& Eckart
2010), both of which can disrupt the circumstellar discs and prematurely
terminate the accretion process, as recently observed in the centre of
the Orion Nebula Cluster (Mann et al. 2014). It is, however, apparent
that there are systematic differences in the accretion process between
different environments, {with the mass accretion process lasting
longer at low metallicity. We note that our conclusions appear to not
agree with claims by Yasui et al. (2009; 2010; 2016) that in some
low-metallicity Galactic environments the lifetimes of discs are instead
shorter. However, those works measure only the apparent dust content of
circumstellar discs and use that as a proxy for the total mass of the
discs and for their lifetimes. We instead measure directly the infall of
the much more abundant gas onto the stars. 

Of course, the conclusions that we have } drawn so far are necessarily
still preliminary, because they are based on a limited sample of star
forming regions in the MW and MCs. Nonetheless, given the accuracy on
the mass accretion rates derived photometrically with the method
discussed in this and previous articles in this series, as the sample of
studied regions grows it will be possible to set stronger constraints on
the dependence of the mass accretion process on the general properties
of the environment. {Even stronger constraints on the accretion
physics, and in particular on the kinematics of the infalling gas, will
come from NIR multi-object spectroscopy of hundreds of PMS objects in
forming regions in the Galaxy and MCs with the James Webb Space
Telescope (Gardner et al. 2007). These guaranteed-time
observations\footnote{The list of approved observations is available at
https://jwst-docs.stsci.edu/display/JSP/JWST+GTO+Observation+Specifications}
will also include coordinated NIR photometry of the targets, providing
for the first time a solid and simultaneous characterisation of the gas
and dust properties of circumstellar discs in these low-metallicity
environments. }

\begin{figure*}[t]
\centering
\resizebox{\hsize}{!}{\includegraphics{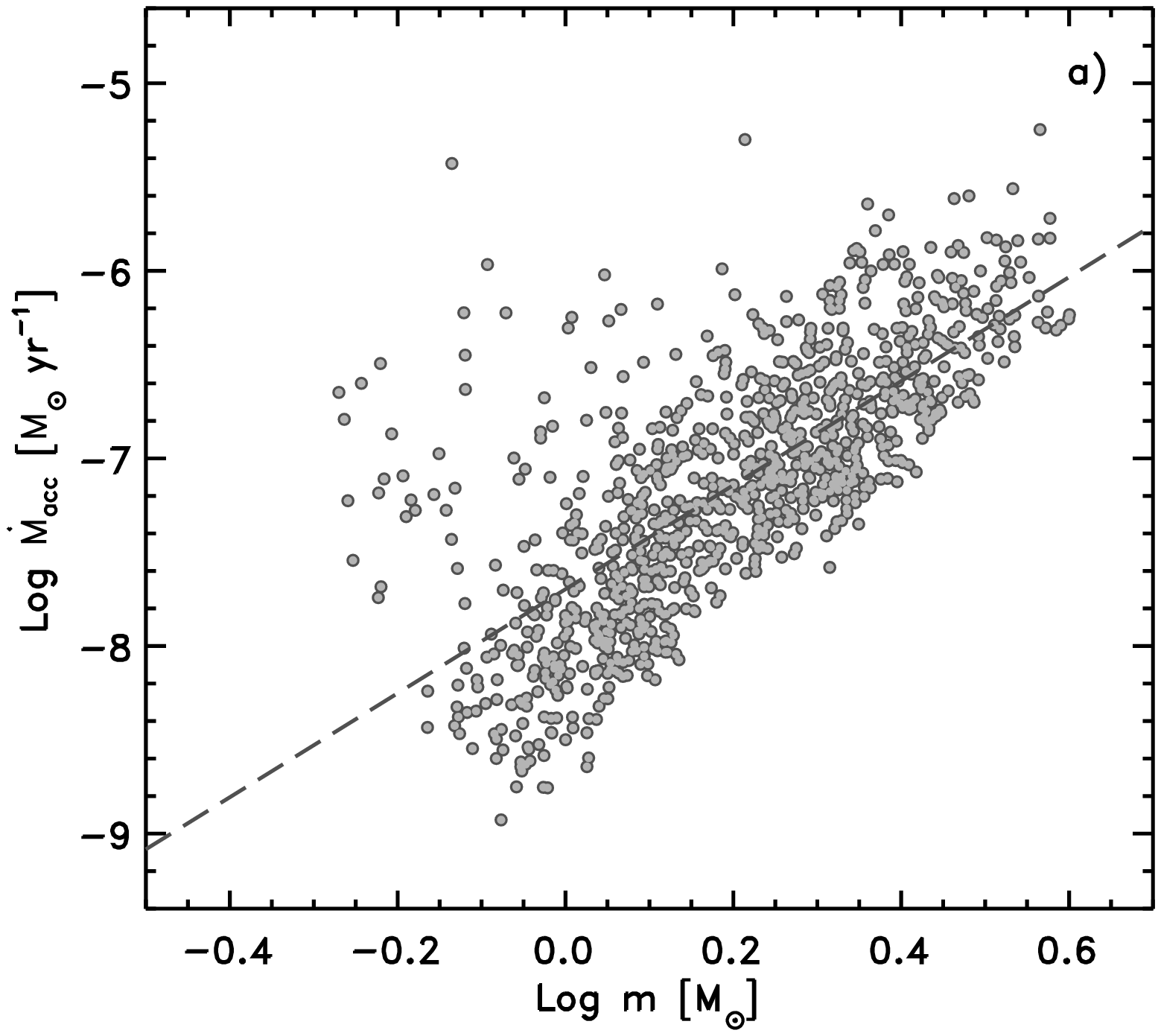}
                      \includegraphics{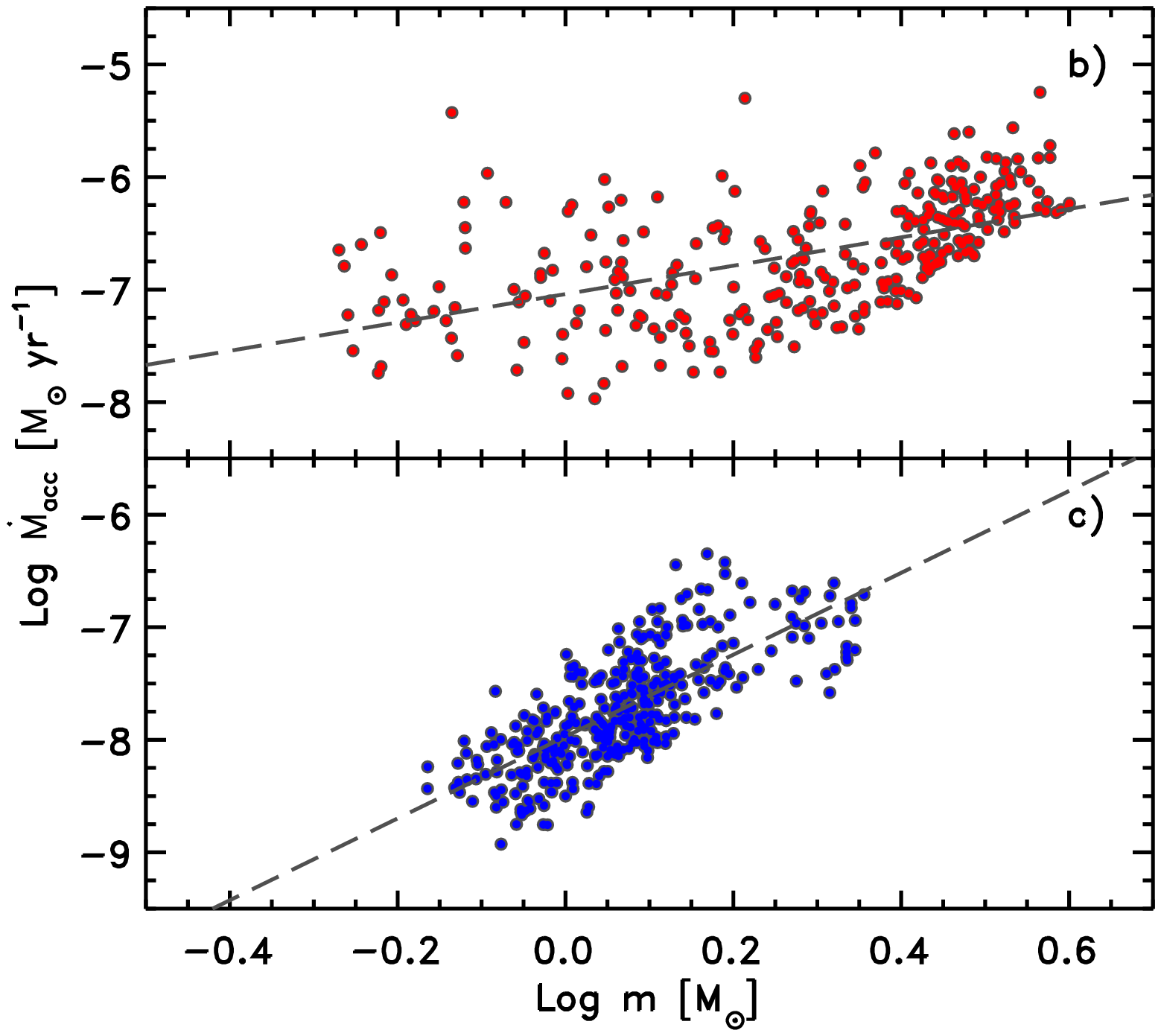}}
\caption{Mass accretion rate of the 30\,Dor bona-fide PMS stars, as a
function of mass. Panel a): all 1\,035 PMS stars. Panel b): PMS objects
with ages between $0.5$ and 2\,Myr. Panel c): PMS stars in the range 
$8 - 16$\,Myr. The dashed lines represent the best fit to the
distribution in each panel.} 
\label{fig12}
\end{figure*}

\begin{acknowledgements}

We are indebted to an anonymous referee, whose constructive criticism
has helped us to improve the presentation of this work. NP acknowledges
partial support by HST-NASA grants GO-11547.06A and GO-11653.12A, and
STScI-DDRF grant D0001.82435.

\end{acknowledgements}

\section*{Appendix}
\subsection*{\small The pitfalls of piecemeal best fitting}

The mass accretion rates measured in 30\,Dor, NGC\,346, and NGC\,602 are
not compatible with a mass dependence as steep as those derived from the
studies of small nearby Galactic star forming regions cited in
Section\,6 (e.g. Muzerolle et al. 2003, 2005; Natta et al. 2004, 2006;
Calvet et al. 2004). Actually, the indices as steep as $m^2$ or more
reported in the literature might be a consequence of having ignored
existing age spreads in the population under study, as well as the
effect of age on $\dot M_{\rm acc}$.

We illustrate these effects in Figure\,\ref{fig12}, where the $\dot
M_{\rm acc}$ values measured in 30\,Dor are displayed as a function of
$m$. We first consider all PMS stars together (Panel a). The formal best
fit to the overall distribution would seem to suggest a relatively steep
dependence of the type $\dot M_{\rm acc} \propto m^\beta$ ({dashed}
line) with $\beta = 2.72 \pm 0.08$. Although formally correct, fitting
the observed snapshot in Panel a) ignores the fact that the ages of
these stars span a range of at least $\sim 30$\,Myr and cannot be
considered coeval and the simultaneous dependence on age cannot be
ignored.


To alleviate this problem one should split the sample into smaller groups
of PMS stars of similar age, for instance all stars within a factor of 2
of a given age. An example is shown in Panel b) for stars around 1\,Myr,
and in this case we find a shallower mass dependence with index $1.22
\pm 0.11$ {(dashed line)}, in agreement with the result of our
multi-variate fit (Section\,6). When different age groups are considered
in the same way, however, the derived slopes will reflect the effects of
evolution. This is what is shown in the example given in Panel c), where
we consider PMS stars with ages within a factor of two of 8\,Myr. The
steeper slope found in this case ($4.2 \pm 0.5$, dashed line) is a
result of several effects, including the general decrease of $\dot
M_{\rm acc}$ with age and the fact that massive PMS objects that have
reached the MS no longer accrete. Since these effects have different
timescales for objects of different masses, and the dependence of the
mass accretion process on mass and age are intertwined, the derived slope
does not have general validity. 

On the other hand, assuming that all stars in our sample formed under
similar conditions and that the functional dependence of the mass 
accretion rate on mass and age does not in turn depend on the value of
these parameters, the multivariate fit discussed in Section\,6
guarantees a robust and meaningful result. It is of course possible
that, for instance, the way in which $\dot M_{\rm acc}$ scales with $t$
is different for different masses. Indeed Figure\,\ref{fig10} reveals
small variations in the value of the slope $\alpha$ between the four
mass groups. However, in spite of these possible limitations, the
adopted power-law description offers a practical way to explore the
parameter space and makes it easier to identify the effects of the
environment on the mass accretion process.


\begin{thebibliography}{}


\bibitem[()]{} Alencar, S., Johns--Krull, C., Basri, G. 2001, AJ,
               122, 3335

\bibitem[()]{} Andersen, M., Zinnecker, H., Moneti, A., McCaughrean, M., 
               Brandl, B., Brandner, W., Meylan, G., Hunter, D. 2009, 
               ApJ, 707, 1347 

\bibitem[()]{} Ardila, D., Herczeg, G., Gregory, S., et al. 2013, ApJS, 
               207, 1 

\bibitem[()]{} Asplund, M., Grevesse, N., Sauval, A. J., Scott, P. 2009, 
               ARA\&A, 47, 481

\bibitem[()]{} Barentsen, G., Vink, J., Drew, J., et al. 2011, MNRAS, 415, 103

\bibitem[()]{} Beccari, G., De Marchi, G., Panagia, N., et al. 2015, A\&A, 
               574, 44

\bibitem[()]{} Beccari, G., Spezzi, L., De Marchi, G., et al. 2010, ApJ, 
               720, 1108 

\bibitem[()]{} Bessell, M., Castelli, F., Plez, B. 1998, A\&A, 333, 231 

\bibitem[()]{} Brandl, B., Sams, B. J., Bertoldi, F., Eckart, A., Genzel, 
               R., Drapatz, S., Hofmann, R., Loewe, M., Quirrenbach, A. 
               1996, ApJ, 466, 254 

\bibitem[()]{} Bressan, A., et al. 2012, MNRAS, 427, 127

\bibitem[()]{} Caffau, E., Ludwig, H., Bonifacio, P., et al. 2010, A\&A, 
               514, A92

\bibitem[()]{} Caffau, E., Ludwig, H., Steffen, M., Freytag, B., 
               Bonifacio, P. 2011, Solar Phys., 268, 255

\bibitem[()]{} Calvet, N., Hartmann, L., Strom, E. 2000, in ``Protostars and
              Planets'', eds V. Mannings, A. Boss, S. Russell (Tucson: 
              University of Arizona Press), 377

\bibitem[()]{} Calvet, N., Muzerolle, J., Brice\~no, C., Hernandez, J.,
               Hartmann, L., Saucedo, J. L., Gordon, K. D. 2004, AJ, 128, 1294

\bibitem[()]{} Carlson, L., Sabbi, E., Sirianni, M., et al. 2007, ApJ, 665, 
               L109 

\bibitem[()]{} Cignoni, M., Sabbi, E., Nota, A., et al. 2009, AJ, 137, 3668


\bibitem[()]{} Cignoni, M., Sabbi, E., van der Marel, R., et al. 2016, 
               ApJ, 833, 154 

\bibitem[()]{} Cusano, F., Ripepi, V., Alcal\'a, J., et al. 2011, MNRAS, 
               410, 227 

\bibitem[()]{} Dahm, S. 2008, AJ, 136, 521

\bibitem[()]{} Da Rio, N., Robberto, M. 2012, AJ, 144, 176

\bibitem[()]{} de Koter, A., Heap, S., Hubeny, I. 1998, ApJ, 509, 879 

\bibitem[()]{} De Marchi, G., Beccari, G, Panagia, N. 2013a, ApJ, 775, 68

\bibitem[()]{} De Marchi, G., Nota, A., Leitherer, C., Ragazzoni, R.,
               Barbieri, C. 1993,  ApJ, 419, 658 

\bibitem[()]{} De Marchi, G., Panagia, N. 2014, MNRAS, 445, 93 

\bibitem[()]{} De Marchi, G., Panagia, N., Girardi, L. 2014, MNRAS, 438, 513 

\bibitem[()]{} De Marchi, G., Panagia, N., Guarcello, M.G., Bonito, R.
               2013b, MNRAS, 435, 3058 

\bibitem[()]{} De Marchi, G., Panagia, N., Romaniello, M. 2010, ApJ, 715, 1 

\bibitem[()]{} De Marchi, G., Panagia, N., Romaniello, M., Sabbi, E.,
               Sirianni, M., Prada Moroni, P., Degl'Innocenti, S. 2011a,
               ApJ, 740, 11 

\bibitem[()]{} De Marchi, G., Panagia, N., Sabbi, E. 2011b, ApJ, 740, 10

\bibitem[()]{} De Marchi, G., Panagia, N., Sabbi, E., et al. 2016, 
               MNRAS, 455, 4373 
               
\bibitem[()]{} De Marchi G., Paresce, F., Panagia, N., et al. 2011c, ApJ, 
               739, 27

\bibitem[()]{} Deustua, S., Mack, J., Bajaj, V., Khandrika, H. 2017, 
               ``Instrument Science Report WFC3 2017-14'' (Baltimore:
               STScI) 

\bibitem[()]{} Dufour, R. 1984, in ``Structure and evolution of the
               Magellanic Clouds'', IAU Symp. 108, eds. S. van den Bergh, A. 
               de Koter (Dordrecht: Reidel), 353

\bibitem[()]{} Durisen, R., Boss, A., Mayer, L., et al. 2007, in 
                 ``Protostars and Planets V'', eds B. Reipurth, D. Jewitt, 
                 K. Keil, (Tucson: University of Arizona Press), 607

\bibitem[()]{} Elson, R., Gilmore, G., Santiago, B. 1997, MNRAS, 289, 157 

\bibitem[()]{} Evans, C., Hainich, R., Oskinova, L., et al. 2012, ApJ, 753, 
               173 

\bibitem[()]{} Fang, M., van Boekel, R., Wang, W., et al. 2009, A\&A, 504, 461 

\bibitem[()]{} Fedele, D., van den Ancker, M., Henning, T., Jayawardhana, R.,
               Oliveira, J. M. 2010, A\&A, 510, A72

\bibitem[()]{} Fernandez, M., Ortiz, E., Eiroa, C., Miranda, L.  1995,
               A\&AS, 114, 439

\bibitem[()]{} Fitzpatrick, E., Savage, B. 1984, ApJ, 279, 578 

\bibitem[()]{} Gardner, J., Mather, J., Clampin, M., et al. 2006, SSRv,
               123, 485

\bibitem[()]{} Geha, M. et al. 1998, AJ, 115, 1045

\bibitem[()]{} Girardi, L., Groenewegen, M., Weiss, A., Salaris, M.
               1998, MNRAS, 301, 149

\bibitem[()]{} Girardi, L., Bertelli, G. 1998, MNRAS, 300, 533

\bibitem[()]{} Gouliermis, D., Schmeja, S., Dolphin, A., et al. 2012, 
               ApJ, 748, 64 

\bibitem[()]{} Grebel, E., Chu, Y.-H. 2000, AJ, 119, 787 

\bibitem[()]{} Gullbring, E., Hartmann, L., Brice\~no, C., Calvet, N. 1998, 
               ApJ, 492, 323

\bibitem[()]{} Habing, H. 1997, in ``Science with the VLT Interferometer'',
               ed. F. Paresce (Berlin: Springer), 183

\bibitem[()]{} Hartmann, L., Calvet, P., Gullbring, E., D'Alessio, P. 1998, 
               ApJ, 495, 385

\bibitem[()]{} Hennekemper, E., Gouliermis, D., Henning, T., Brandner, W.,
               Dolphin, A. 2008, ApJ, 672, 914

\bibitem[()]{} Hill, V., Andrievsky, S., Spite, M. 1995, A\&A, 293, 347

\bibitem[()]{} Hunter, D., O'Neil, E., Lynds, R., Shaya, E., Groth, E., 
               Holtzman, J. 1996, ApJ, 459, L27 

\bibitem[()]{} Hunter, D., Shaya, E., Holtzman, J., Light, R., O'Neil, E., 
               Lynds, R. 1995, ApJ, 448, 179 

\bibitem[()]{} Hunter, D., Shaya, E., Scowen, P., Hester, J., Groth, E., 
               Lynds, R., O'Neil, E. 1995b, ApJ, 444, 758

\bibitem[()]{} Hur, H., Sung, H., Bessell, M. 2012, AJ, 143, 41 

\bibitem[()]{} Ingleby, L., Calvet, N., Herczeg, G., et al. 2013, ApJ, 767, 
               112 

\bibitem[()]{} Kalari, V., Vink, J. 2015, ApJ, 800, 113

\bibitem[()]{} K\"onigl, A. 1991, ApJ, 370, L39

\bibitem[()]{} Lee, J.-K., Rolleston, W., Dufton, P., Ryans, R. 2005,
               A\&A, 429, 1025 

\bibitem[()]{} Leitherer, C., Schaerer, D., Goldader, J., et al. 1999, 
               ApJS, 123, 3 

\bibitem[()]{} Lilly, S., Le Fevre, O., Hammer, F., Crampton, D. 1996, ApJ,
               460, L1

\bibitem[()]{} Lodders, K. 2010, in ``Principles and Perspectives in 
               Cosmochemistry'', ASSP, (Berlin: Springer), 379

\bibitem[()]{} Lortet, M., Testor, G. 1991, A\&AS, 89, 185 

\bibitem[()]{} Madau, P., Ferguson, H., Dickinson, M., Giavalisco, M., et al.
                 1996, MNRAS, 283, 1388


\bibitem[()]{} Mann, R., Di Francesco, J., Johnstone, D., et al. 2014, ApJ, 
               784, 82 

\bibitem[()]{} Marigo, P., Girardi, L., Bressan, A., Groenewegen, M., 
               Silva, L., Granato, G. 2008, A\&A, 482, 883 

\bibitem[()]{} Massey, P., Hunter, D. 1998, ApJ, 493, 180 

\bibitem[()]{} Melnick, J. 1987, in ``Observational Evidence of Activity 
               in Galaxies'', IAU Symp. 121, Ed. E. Khachikian, K.
               Fricke, J. Melnick (Dordrecht: Reidel), 545

\bibitem[()]{} Meynet, G., Maeder, A., Schaller, G., Schaerer, D., 
               Charbonnel, C. 1994, A\&AS, 103, 97

\bibitem[()]{} Muzerolle, J., Calvet, N., Brice\~no, C., Hartmann, L.,
               Hillenbrand, L. 2000, ApJ, 535, L47

\bibitem[()]{} Muzerolle, J., Hillenbrand, L., Calvet, N., Brice\~no, C.,
               Hartmann, L., 2003, ApJ, 592, 266

\bibitem[()]{} Muzerolle, J., Luhman, K., Brice\~no, C., Hartmann, L.,
               Calvet, N. 2005, ApJ, 625, 906

\bibitem[()]{} Natta, A., Testi, L., Muzerolle, J., Randich, S., et al.
               2004, A\&A, 424, 603

\bibitem[()]{} Natta, A., Testi, L., Randich, S. 2006, A\&A, 452, 245

\bibitem[()]{} Nigra, L., Gallagher, J., Smith, L., et al. 2008, PASP, 120, 
               972 

\bibitem[()]{} Olczak, C., Pfalzner, S., Eckart, A. 2010, A\&A, 509, A63 

\bibitem[()]{} Oskinova, L., Sun, W., Evans, C., et al. 2013, ApJ, 765, 73

\bibitem[()]{} Panagia, N. 1973, AJ, 78, 929

\bibitem[()]{} Panagia, N. 1999, in ``New Views of the Magellanic
               Clouds'', IAU Symp. 190, eds. Y.-H. Chu, N. Suntzeff, J. 
               Hesser, D. Bohlender (San Francisco: ASP), 549

\bibitem[()]{} Panagia, N., Gilmozzi, R., Macchetto, F., Adorf, H.-M.,
               Kirshner, R. P.  1991, ApJ, 380, L23

\bibitem[()]{} Romaniello, M. 1998, PhD thesis, Scuola Normale Superiore,
               Pisa, Italy

\bibitem[()]{} Romaniello, M., Panagia, N., Scuderi, S., Kirshner, R. P.
               2002, AJ, 123, 915 

\bibitem[()]{} Russell, S., Dopita, M. 1990, ApJS, 74, 93 

\bibitem[()]{} Sabbi, E., Sirianni, M., Nota, A., et al. 2007, AJ, 133, 44

\bibitem[()]{} Sana, H., Momany, Y., Gieles, M., et al. 2010, A\&A, 515, A26
 
\bibitem[()]{} Schmalzl, M., Gouliermis, D., Dolphin, A., Henning, T. 2008, 
               ApJ, 681, 290

\bibitem[()]{} Selman, F., Melnick, J., Bosch, G., Terlevich, R. 1999,
               A\&A, 347, 532 

\bibitem[()]{} Shu, F., Najita, J., Ruden, S., Lizano, S. 1994, ApJ, 429, 797

\bibitem[()]{} Sicilia--Aguilar, A., Hartmann, L., Furesz, G., et al. 2006, 
               AJ, 132, 2135

\bibitem[()]{} Sicilia-Aguilar, A., Henning, T., Hartmann, L. 2010, ApJ,
               710, 597

\bibitem[()]{} Sirianni, M., Nota, A., Leitherer, C., De Marchi, G., 
               Clampin, M. 2000, ApJ, 533, 203 

\bibitem[()]{} Smartt, S., Lennon, D., Kudritzki, R., et al. 2002, A\&A, 
               391, 979 

\bibitem[()]{} Smith, K., Lewis, G., Bonnell, I., Bunclark, P., Emerson,
               J. 1999, MNRAS, 304, 367

\bibitem[()]{} Spezzi, L., De Marchi, G., Panagia, N., et al. 2012, MNRAS, 
               421, 78

\bibitem[()]{} Tognelli, E., Prada Moroni, P., Degl'Innocenti, S. 2012,
               A\&A, 533, A109

\bibitem[()]{} Valenti, J., Basri, G., Johns, C. 1993, AJ, 106, 2024

\bibitem[()]{} Vassiliadis, E., Wood, P. 1994, ApJS, 92, 125

\bibitem[()]{} von Braun, K., Chiboucas, K., Minske, J. K., Salgado, J., 
               Worthey, G.  1998, PASP, 110, 810

\bibitem[()]{} Walborn, N. 1986, in ``Luminous Stars and Associations 
               in Galaxies'', IAU Symp. 116, Ed. W. de Loore, A. Willis,
               P. Laskarides (Dordrecht: Reidel), 185 

\bibitem[()]{} Walborn, N. 1991, in ``The Magellanic Clouds'', IAU Symp.
               148, Ed. R. Haynes, D. Milne (Dordrecht: Kluwer), 145

\bibitem[()]{} Walborn, N., Barba, R., Brandner, W., Rubio, M., Grebel, 
               E., Probst, R. 1999, AJ, 117, 225 

\bibitem[()]{} Walborn, N., Blades, J. 1997, ApJS, 112, 457 

\bibitem[()]{} Westerlund, B., Linde, P., Lyng\aa, G. 1995, A\&A, 298, 39 

\bibitem[()]{} Yasui, C., Kobayashi, N., Tokunaga, A., Saito, M., Tokoku, 
               C. 2009, ApJ, 705, 54 

\bibitem[()]{} Yasui, C., Kobayashi, N., Tokunaga, A., Saito, M., Tokoku, 
               C. 2010, ApJ, 723, L113 

\bibitem[()]{} Yasui, C., Kobayashi, N., Tokunaga, A., Saito, M., Izumi, 
               N. 2016, AJ, 151, 50 

\bibitem[()]{} Zeidler, P., Grebel, E., Nota, A., et al. 2016, AJ, 152, 84 

\bibitem[()]{} Zinnecker, H., Andersen, M., Brandl, B., Brandner, W.,
               Hunter, D., Larson, R., McCaughrean, M., Meylan, G., 
               Moneti, A. 2002, in ``Extragalactic Star Clusters,'' IAU 
               Symp. 207, eds. D. Geisler, E. Grebel, D. Minniti (San 
               Francisco: ASP), 531 



\end{thebibliography}
\end{document}